\documentclass[twocolumn]{aastex7}
\usepackage{amsmath}

\newcommand{\lrb}[1]{\left({#1}\right)}

\begin{document}

\title{First Light of Neutron Star Mergers: Off-axis Cocoon Cooling X-ray Emission from Short Gamma-Ray Burst Jets}

\author[orcid=0000-0001-5751-633X]{Jian-He Zheng} 
\affiliation{School of Astronomy and Space Science, Nanjing University, Nanjing 210023, People’s Republic of China}
\affiliation{Department of Astronomy and Theoretical Astrophysics Center, University of California, Berkeley, CA 94720, USA}
\email[show]{mg21260020@smail.nju.edu.cn}  

\author[0000-0002-1568-7461]{Wenbin Lu}
\affiliation{Department of Astronomy and Theoretical Astrophysics Center, University of California, Berkeley, CA 94720, USA}
\email{wenbinlu@berkeley.com}  

\begin{abstract}
Neutron star mergers (NSMs) are confirmed gravitational wave sources. Identifying an early electromagnetic counterpart for these events is therefore crucial for rapid localization and multimessenger follow-up. However, the associated gamma-ray bursts (GRBs) are highly collimated and are therefore easily missed by off-axis observers. An early, less beamed counterpart is essential for identifying the majority of mergers.
In this Letter, we investigate the cooling emission from jet-driven cocoons produced by short gamma-ray burst jets propagating through merger ejecta. We perform hydrodynamic simulations and radiative post-processing to calculate the early X-ray emission over a wide range of viewing angles.
We find that the mildly relativistic cocoon produces bright soft X-ray transients for off-axis observers, with luminosities of $10^{46-48}{\rm erg\,s^{-1}}$ and durations of a few to ten seconds. The X-ray spectra are quasi-thermal with characteristic temperatures of $0.1$--$1\,{\rm keV}$. 
For observers at $\theta_{\rm v}=10^{\circ}$, the cocoon emission is detectable by Einstein Probe (EP) out to $z\simeq 0.3$. For nearby events like GW170817, it remains detectable up to $\theta_{\rm v}\simeq 45^{\circ}$. 
The predicted detection rate for EP is $0.5\,{\rm yr^{-1}}$ in the canonical model.
In future multimessenger campaigns, rapid UV/optical/IR follow-up of such X-ray triggers can subsequently identify the associated kilonova and jet afterglow emission, which can confirm their origin.

\end{abstract}


\section{Introduction} 
Neutron star mergers (NSMs) are one of the most important targets in multimessenger astronomy. They generate gravitational waves (GWs), eject neutron-rich material that powers kilonova emission, and launch relativistic jets that produce gamma-ray bursts (GRBs) \citep{Paczynski1986ApJ...308L..43P,Eichler1989Natur.340..126E,Li1998ApJ...507L..59L,Metzger2010MNRAS.406.2650M,Metzger2020LRR....23....1M}. 
The discovery of GW170817 validates this multimessenger picture. It was accompanied by the weak short gamma-ray burst (SGRB) GRB 170817A and the optical/infrared kilonova AT2017gfo \citep{GW170817PhRvL.119p1101A,GW1708172017ApJ...848L..13A,GRB170817_2017ApJ...848L..14G,Coulter2017Sci...358.1556C, 2017Sci...358.1570D, Shappee2017Sci...358.1574S,Kasen2017Natur.551...80K,Smartt2017Natur.551...75S}. 

GW170817 was an exceptionally nearby event. Thanks to its proximity, the early kilonova was identified rapidly, allowing intensive follow-up observations \citep{2021ARA&A..59..155M}. 
For more distant NSMs, the kilonova will be much fainter and difficult to identify via optical surveys. In these cases, early high-energy signals become essential for rapid follow-up and counterpart identification \citep{Metzger2012ApJ...746...48M}.

Short GRBs provide the brightest high-energy signal of NSMs, but their gamma-rays are highly collimated and detectable only for observers close to the jet axis \citep{Berger2014ARA&A..52...43B}. Most NSMs have jet axes far from our line of sight, so GRBs would be missed. Therefore, the key challenge is to identify an early, bright, and less-beamed counterpart that can trigger rapid follow-up for the majority of NSMs.

The launch of Einstein Probe (EP) provides a new opportunity to search for such signals. Its Wide-field X-ray Telescope (WXT) is designed to monitor a large fraction of the sky in the soft X-ray band ($0.5$--$4{\rm keV}$) \citep{Yuan2022hxga.book...86Y,Yuan2025SCPMA..6839501Y}.
Recent EP observations have detected prompt soft X-ray emission from both long and short GRBs, demonstrating that this band can reveal components of relativistic explosions \citep[e.g.,][]{Liu2025NatAs...9..564L,Levan2025NatAs...9.1375L,Li2026_250704B}.
Meanwhile, EP has discovered several Fast X-ray Transients (FXTs) associated with broad-lined Type Ic supernovae (SNe Ic-BL) but without gamma-ray counterparts \citep[e.g.,][]{Sun2025NatAs...9.1073S,Li2025arXiv250417034L}. These events suggest that soft X-ray surveys can uncover jet-driven explosions with weak gamma-ray emission.

The origins of gamma-ray-dark FXTs remain uncertain. One possibility is an on-axis but intrinsically weak jet, whose Lorentz factor or energy is too low to produce gamma-ray emission \citep[e.g.,][]{Sun2025NatAs...9.1073S,Hamidani2025ApJ...986L...4H}. Another possibility is an ordinary GRB jet viewed off-axis, for which the prompt gamma-rays are beamed away from the line of sight \citep[e.g.,][]{Yamazaki2002ApJ...571L..31Y,Zheng2025ApJ...985...21Z}. We focus on the second scenario: ordinary short-GRB jets launched in NSMs, propagating through pre-existing merger ejecta, and producing mildly relativistic cocoons whose cooling X-ray emission may be detectable even when the gamma rays are missed.

A relativistic jet launched after an NSM must propagate through the surrounding ejecta before breakout. During this process, shocks at the jet head deposit energy into the ambient material and create a hot cocoon around the jet, similar to cocoon formation in long GRBs inside stellar envelopes \citep{Nagakura2014ApJ...784L..28N,Lazzati2017ApJ...848L...6L,Gottlieb2018Cocoon,Hamidani2021Propagation}.  Previous hydrodynamic studies examined jet collimation and breakout through merger ejecta \citep{Nagakura2014ApJ...784L..28N,Duffell2018ApJ...866....3D}. In particular, \citet{Duffell2018ApJ...866....3D} showed that powerful, short GRB-like jets are difficult to choke. However, these studies focused on the jet--ejecta hydrodynamics and did not follow the long-term cocoon evolution or calculate the cocoon emission.

The existence of such cocoon material is also observationally supported by GRB 170817A, whose prompt emission showed a hard main pulse followed by a softer thermal tail \citep{GRB170817_2017ApJ...848L..14G,Zhang2018GRB170817}. This component has been interpreted as cocoon shock breakout emission, which can explain its unusually low luminosity and the deviation from the Amati relation \citep{Gottlieb2018CocoonSBO}.


Cocoon radiation contains two components: shock-breakout emission and subsequent cocoon cooling emission \citep[e.g.,][]{Nakar2020PhR...886....1N}. Shock breakout occurs when radiation trapped near the shock begins to diffuse through the outer ejecta, typically at relatively small radii ($R_{\rm bo}\sim 10^{10}-10^{11}$cm). It produces a short flash that can extend to hard X-rays or gamma-rays, particularly when the breakout layer is out of local thermodynamic equilibrium \citep{Gottlieb2018CocoonSBO,gutierrez2025PhRvD.111f3031G,Pais2026arXiv260629515P}. 
After breakout, the bulk shock-heated cocoon continues to expand, and its stored thermal energy is released at much larger radii through cooling and photon diffusion, producing a longer and softer signal. 
Cocoon cooling emission from an NSM has not yet been identified in X-rays because of the lack of an EP-like wide-field soft X-ray monitor during GW170817. Now EP is discovering gamma-ray-dark FXTs whose origins are not clear. FXTs from NSM cocoon cooling could represent an unrecognized population among these events. This Letter focuses on predicting this soft X-ray emission.

Cocoon cooling emission from NSMs has been studied in previous analytical and numerical works \citep[e.g.,][]{Nakar2017cocoon,Lazzati2017ApJ...848L...6L,Gottlieb2018Cocoon,Wang2018EarlyXray,Hamidani2023Cooling}. 
Previous studies did not provide detailed viewing-angle-dependent soft X-ray lightcurves during $t_{\rm obs}<10$ s, when the X-ray is expected to peak and rapidly fade.
They generally adopted semi-analytical models or used early-time hydrodynamic simulations ($t_{\rm lab}\sim10$ s) followed by analytical or semi-analytical calculations of the later cooling emission.
These approaches have clarified the basic physics, but they do not fully follow the long-term evolution { ($t_{\rm lab}>10^3$ s)} of the cocoon to the photon diffusion radius, where the observed cooling emission is released. Consequently, modeling the observable signal requires extrapolating the cocoon evolution beyond simulation time.

{ The main novelty of this Letter is to directly connect the long-term cocoon evolution to its observable soft X-ray signal.} We simulate the jet-cocoon evolution {\rm directly} to the diffusion radius and apply radiative post-processing to calculate the early, viewing-angle-dependent X-ray lightcurves and spectra from the simulated structure. { We further quantify the luminosity function and expected detection rate in the EP-WXT band, making cocoon cooling a testable channel for identifying off-axis NSMs.}

The paper is organized as follows: In Section \ref{sec:methods}, we describe our simulation setup. Then, we show X-ray lightcurves of cooling emission from the cocoon in Section \ref{sec:res}. The detection rate and verification strategy are in Sections \ref{sec:rate} and \ref{sec:veri}.  Throughout this paper, we use CGS units and ignore cosmological redshift factors (i.e., the observable quantities are expressed in the host galaxy's rest frame).

\begin{figure*}
    \centering
    \includegraphics[width=\textwidth]{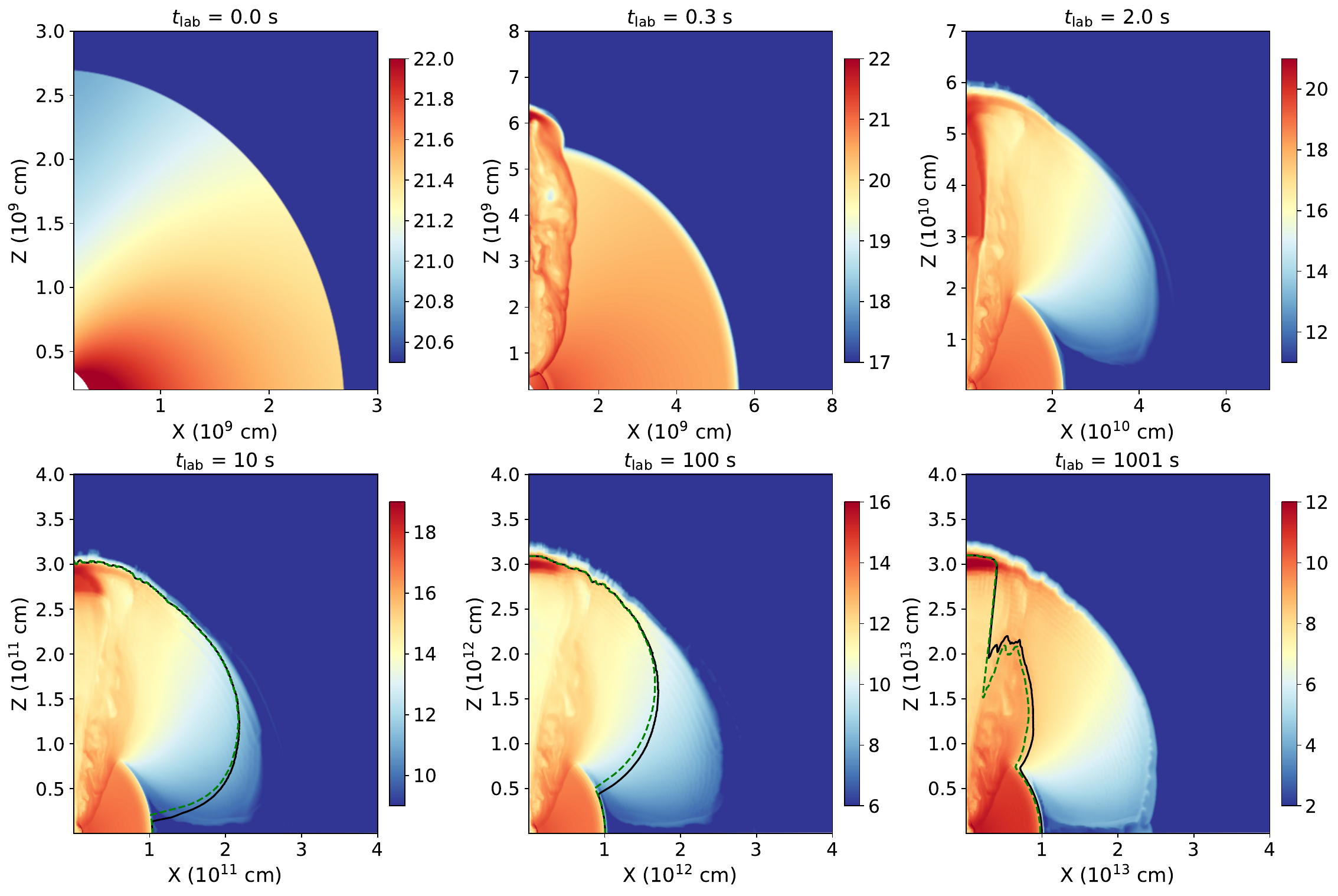}
    \caption{
    Energy density ($\Gamma(\Gamma-1)\rho'c^2+\Gamma^2 e'$) maps of the canonical model (Lc). 
    Six panels correspond to different lab-frame times.
    The colorbar shows the logarithmic energy density in units of ${\rm erg\,{cm}^{-3}}$. 
    The black solid lines and green dashed lines in the bottom three panels indicate the photospheric radius $r_{\rm ph}$ and diffusion radius $r_{\rm diff}$.
    }
    \label{fig:Engden}
\end{figure*}

\section{Initial Setups}
\label{sec:methods}
Our simulations are designed to capture the jet-ejecta interaction from the NSM ejecta to the cocoon diffusion radius, where the photon diffusion time becomes comparable to the dynamical expansion time. In the NSM scenario, the cocoon diffusion radius is of order $\sim10^{13}\,{\rm cm}$, corresponding to a lab-frame time of $\sim10^{3}\,{\rm s}$. This radius is smaller than that in the collapsar scenario because the NSM ejecta and its corresponding cocoon are less massive.
Nevertheless, resolving this long-term evolution in full three dimensions is computationally demanding. We therefore carry out two-dimensional axisymmetric relativistic hydrodynamic simulations using the public code \texttt{PLUTO} v4.4 \citep{Mignone2007ApJS..170..228M}. This approach enables us to investigate different jet and ejecta parameters efficiently.

The properties of the NSM ejecta before jet launching remain uncertain and depend on the binary masses, neutron star equation of state, and merger remnant evolution (see \cite{Shibata2019Review} for a review). For the jet-ejecta interaction, the key quantities are the total (pre-existing) ejecta mass $M_{\rm ej}$, and the delay between the merger and jet launch $t_{\rm delay}$. 
Numerical simulations also show that the ejecta are anisotropic, with more mass concentrated on the equatorial plane \citep[e.g.,][]{Hotokezaka2013PhRvD..87b4001H,Radice2018ApJ...869..130R}. Therefore, we adopt a power-law radial density profile together with an equatorially enhanced angular distribution:

\begin{equation}
\begin{aligned}
\rho_{\rm ej}(r_{\rm min}<r<r_{\rm max},\theta)
&=\rho_0 r^{-n}(0.3+\sin^2\theta) \\
v_{\rm ej}(r_{\rm min}<r<r_{\rm max})&=v_{\rm min}\frac{r}{r_{\rm min}},
\end{aligned}
\end{equation}
where $r_{\rm min}$ and $r_{\rm max}$ are the inner and outer radii of the merger ejecta, respectively. The angular distribution $\rho(\theta)\propto\sin^2\theta$ is motivated by the work of \citet{PeregoAT2017gfo_2017ApJ...850L..37P} as constrained by the observation of kilonova AT2017gfo.
The normalization constant $\rho_0$ is obtained by the total ejecta mass $M_{\rm ej}=\int\rho_{\rm ej}(r,\theta)dV_{\rm ej}$. We assume the ejecta expand homologously ($v_{\rm ej}\propto r$) before the jet launch. The ejecta boundaries are therefore given by
\begin{equation}
    r_{\rm min}=v_{\rm min}t_{\rm delay}, \ \ r_{\rm max}=v_{\rm max}t_{\rm delay},
\end{equation}
where $t_{\rm delay}$ is the delay time between the merger and jet launch. 

The minimum velocity is estimated from the escape velocity of the remnant at $r_{\rm min}$, which means
\begin{equation}
    v_{\rm min}\simeq \lrb{2GM_{\rm rem}\over t_{\rm delay}}^{1/3} \approx 0.03c \lrb{M_{\rm rem}\over 2.7M_\odot}^{1/3} \lrb{t_{\rm delay}\over 1\rm\,s}^{-1/3}.
\end{equation}
We fix $M_{\rm rem}=2.7M_{\odot}$ in our simulations, as our results are only weakly affected by this parameter. Lightcurve modeling of AT2017gfo suggests that the fastest kilonova ejecta reached velocities of order $\sim0.3c$ \citep[e.g.,][]{Kasen2017Natur.551...80K,Nicholl2017ApJ...848L..18N,Shappee2017Sci...358.1574S}; therefore, we adopt $v_{\rm max}=0.3c$. 

{ 
Recent simulations find that a small amount of dynamical ejecta can extend to mildly relativistic velocities, $v\geq0.3c$ \citep[e.g.,][]{Hotokezaka2018ApJ...867...95H,Radice2018Viscous,Rosswog2025MNRAS.538..907R}. We do not include this fast ejecta tail because it is expected to be optically thin by the later cocoon-cooling phase, so it mainly affects the earlier shock-breakout signal rather than the cooling emission considered here.}

The radial density profile is described by a power law $\rho\propto r^{-n}$. The exact value of the power-law index remains uncertain. Numerical simulations generally find steep profiles with $n\sim3$--$4$ in the polar direction \citep{Hotokezaka2013PhRvD..87b4001H,Nagakura2014ApJ...784L..28N}, while shallower distributions with $n\approx2$ can arise when the ejecta are dominated by post-merger disk winds \citep{Fujibayashi2018ApJ...860...64F}. In this paper, we adopt $n=3$. We explored a shallower profile $n=2$ and found that it does not qualitatively change our conclusions.

Observations of AT2017gfo constrain the ejecta mass to be of order $\sim10^{-2}\,M_{\odot}$ \citep[e.g.,][]{PeregoAT2017gfo_2017ApJ...850L..37P,2017Sci...358.1570D,Metzger2020LRR....23....1M,2021ARA&A..59..155M}. Therefore we adopt $M_{\rm ej}\approx10^{-2}\,M_{\odot}$ in the canonical model and explore a broader range of $M_{\rm ej}=3\times10^{-3}$ to $3\times10^{-2}\,M_{\odot}$ to account for the uncertainty in NSM ejecta masses. The parameters of all models are summarized in Table~\ref{tab:param}.

Outside the ejecta, we use a static interstellar medium (ISM) with a smooth density transition 
\begin{equation}
    \rho (r>r_{\rm max})=10^{-8}\rho(r_{\rm max},0^{\circ})e^{-r/r_{\rm max}}+\rho_{\rm ISM}.
\end{equation}
To avoid numerical errors, we adopt the ISM density as $\rho_{\rm ISM}=10^{-18} {\rm g\,cm^{-3}}$. Although this value is higher than a typical ISM density, it remains negligible compared with the densities of the ejecta and cocoon. Consequently, our results are insensitive to the exact choice of $\rho_{\rm ISM}$.

For the pressure profile, we consider that the NSM ejecta are heated by the radioactive decay of freshly synthesized $r$-process nuclei and set $p = 10^{-4}\rho c^2$.
This choice is motivated by the characteristic energy release of order MeV per nucleon from $r$-process nucleosynthesis, corresponding to an energy of $\sim10^{-3}m_{\rm p}c^2$. Assuming the gas is radiation-dominated, the corresponding pressure is $p=e_{\rm int}/3\sim3\times10^{-4}\rho c^2$.
Since only a fraction of the ejecta undergoes strong $r$-process nucleosynthesis, the average thermal energy density of the entire ejecta is expected to be lower. We therefore adopt a conservative value of $p = 10^{-4}\rho c^2$. 
Radioactive heating during the early cocoon-cooling phase is negligible because the cocoon only has a small amount of mass $\sim10^{-4}\,M_{\odot}$, with the corresponding heating rate $\dot\epsilon_r\approx10^{44}\,M_{\rm c,-4}t^{-1.3}_{\rm obs,1}{\rm erg\,s^{-1}}$ much smaller than the cooling luminosity.
For the ambient medium pressure, we adopt $p = 10^{-6}\rho c^2$. We have verified that our results are insensitive to the adopted ambient pressure, since the flow dynamics are dominated by the rest-mass and kinetic energy rather than by the thermal pressure.

\begin{table}[]
    \centering
    \caption{Simulation Model Parameters}
    \begin{tabular}{c|c|c|c|c}
    \hline
        Model & $L_{\rm j}[{\rm erg\,s^{-1}}]$  & $t_{\rm delay}[{\rm s}]$  & $M_{\rm ej}[{ M_{\odot}}]$  & $t_{\rm bo}[{\rm s}]$  \\
    \hline
        Canonical ($Lc$)  &      $10^{50}$     &   0.3  & $10^{-2}$  & 0.3 \\
        Short delay ($Lsd$) &     $10^{50}$    &   0.1  & $10^{-2}$  & 0.2\\
        Long delay ($Lld$) &      $10^{50}$    &   1.0     & $10^{-2}$ & 0.7\\
        Low Mass ($Llm$) &       $10^{50}$     &   0.3  & $3\times10^{-3}$ & 0.2\\
        High Mass ($Lhm$)  &      $10^{50}$    &   0.3  & $3\times10^{-2}$  & 0.4\\
        Powerful ($Lp$) &      $10^{51}$      &   0.3   & $10^{-2}$   &0.2 \\
    \hline
    \end{tabular}
    \tablecomments{
    $L_{\rm j}$ is the two-sided jet luminosity, $t_{\rm delay}$ is the delay time between the merger and jet launch, $t_{\rm bo}$ is the jet breakout time (numerically measured), and the jet duration is fixed to be $t_{\rm j}=1\,$s for all models.
    }
    \label{tab:param}
\end{table}

All simulations are performed in spherical coordinates ($r,\theta$). We adopt the HLL Riemann solver, third-order Runge–Kutta time integration, and piecewise parabolic reconstruction.
The equation of state is polytropic with an adiabatic index of $4/3$, appropriate for the radiation-dominated jet-cocoon system.

The radial grid is divided into two logarithmically uniform patches. The inner patch extends from the inner boundary $r_{\rm min}$ to $6\times10^{11}\,$cm, containing 2000 cells. The outer patch covers the range from $6\times10^{11}\,$cm to $2\times10^{14}\,$cm with 1500 cells. 
The angular grid is a uniform patch, consisting of 400 cells. The full mesh grid therefore consists of $3500\times400$ cells in $r\times\theta$. This corresponds to an angular resolution $\delta\theta=4\times10^{-3}$ rad and a comparable radial resolution $\delta r/r\simeq4\times10^{-3}$. We have verified the numerical convergence in our setups by increasing both angular and radial resolution by a factor of two.

\begin{figure}
    \centering
    \includegraphics[width=\linewidth]{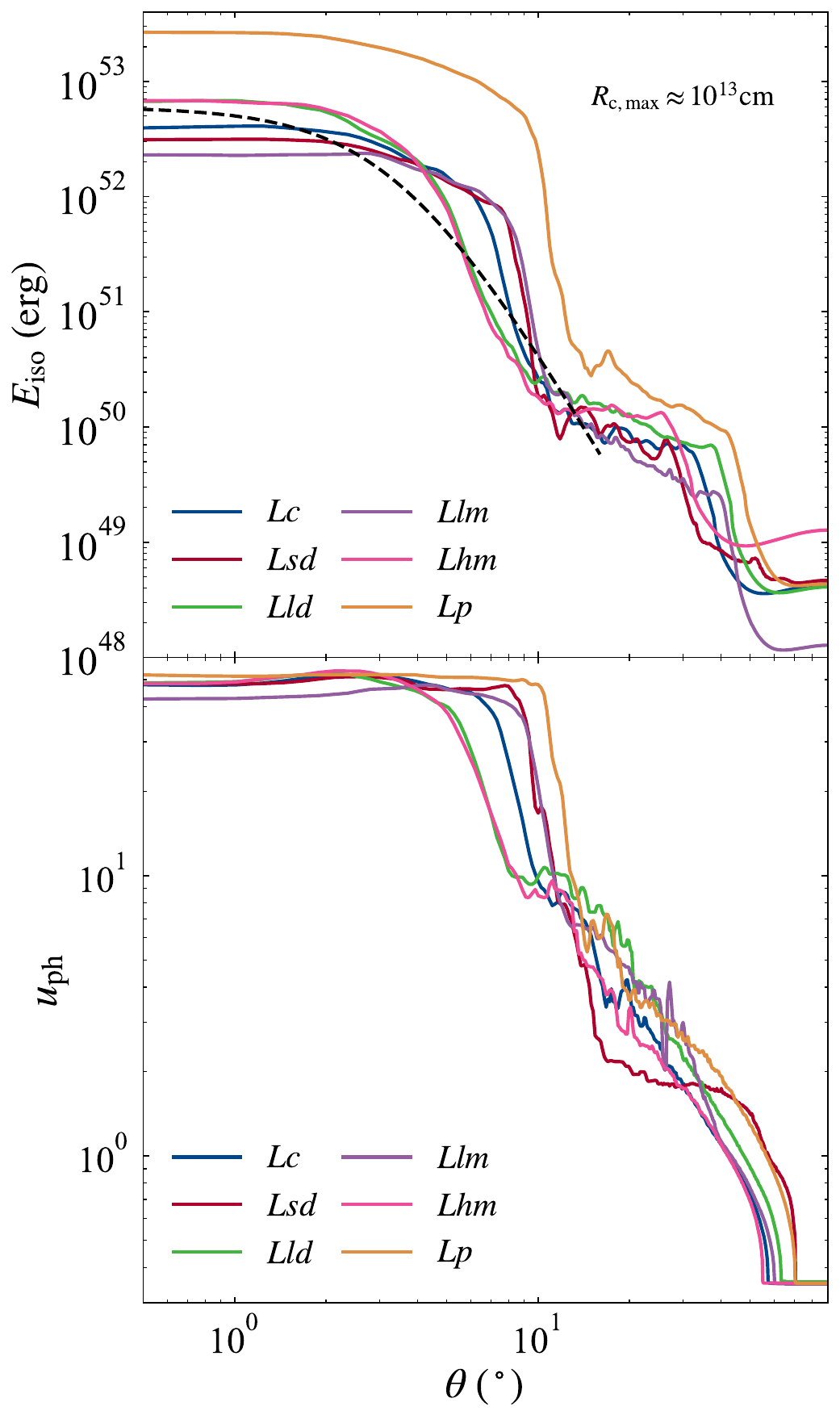}
    \caption{
    \textit{Upper panel:} Angular profiles of the isotropic-equivalent energy $E_{\rm iso}$ for all models at $R_{\rm c,max}\approx10^{13}\,{\rm cm}$. The blue, red, green, purple, pink, and brown curves correspond to models $Lc$, $Lsd$, $Lld$, $Llm$, $Lhm$, and $Lp$, respectively. The black dashed line shows the angular profile used in our afterglow calculations. \textit{Lower panel:} Angular profiles of the photospheric four-velocity, $u_{\rm ph}$, for the same models.
    }
    \label{fig:angular}
\end{figure}

The jet is injected through a nozzle at $r_{\rm min}$ with prescribed radial velocity, density, and enthalpy profiles. Its density and pressure profiles follow ${\rm cosh^{-8}(\theta/\theta_{\rm j,0})}$, producing a nearly uniform core within the half-opening angle $\theta_{\rm j,0}=5^{\circ}$ and a steep decline outside it. The jet is initially hot and launched with a Lorentz factor $\Gamma_0 = 0.7 \theta_{\rm j,0}^{-1}$. 
The maximum terminal four-velocity of the injected jet material is set to $u_{\infty}=50$. This value is lower than that expected for the core of a canonical GRB jet {but offers better numerical stability while being sufficiently high for our purpose in this work, because our observables are largely determined by the cocoon region away from the jet axis at $\theta\gtrsim10^{\circ}$}, rather than the jet itself. We adopt outflow boundary conditions at the outer radial boundary and reflective boundary conditions at both the polar axis and the equatorial plane. 

\begin{figure*}
    \centering
    \includegraphics[width=0.32\textwidth]{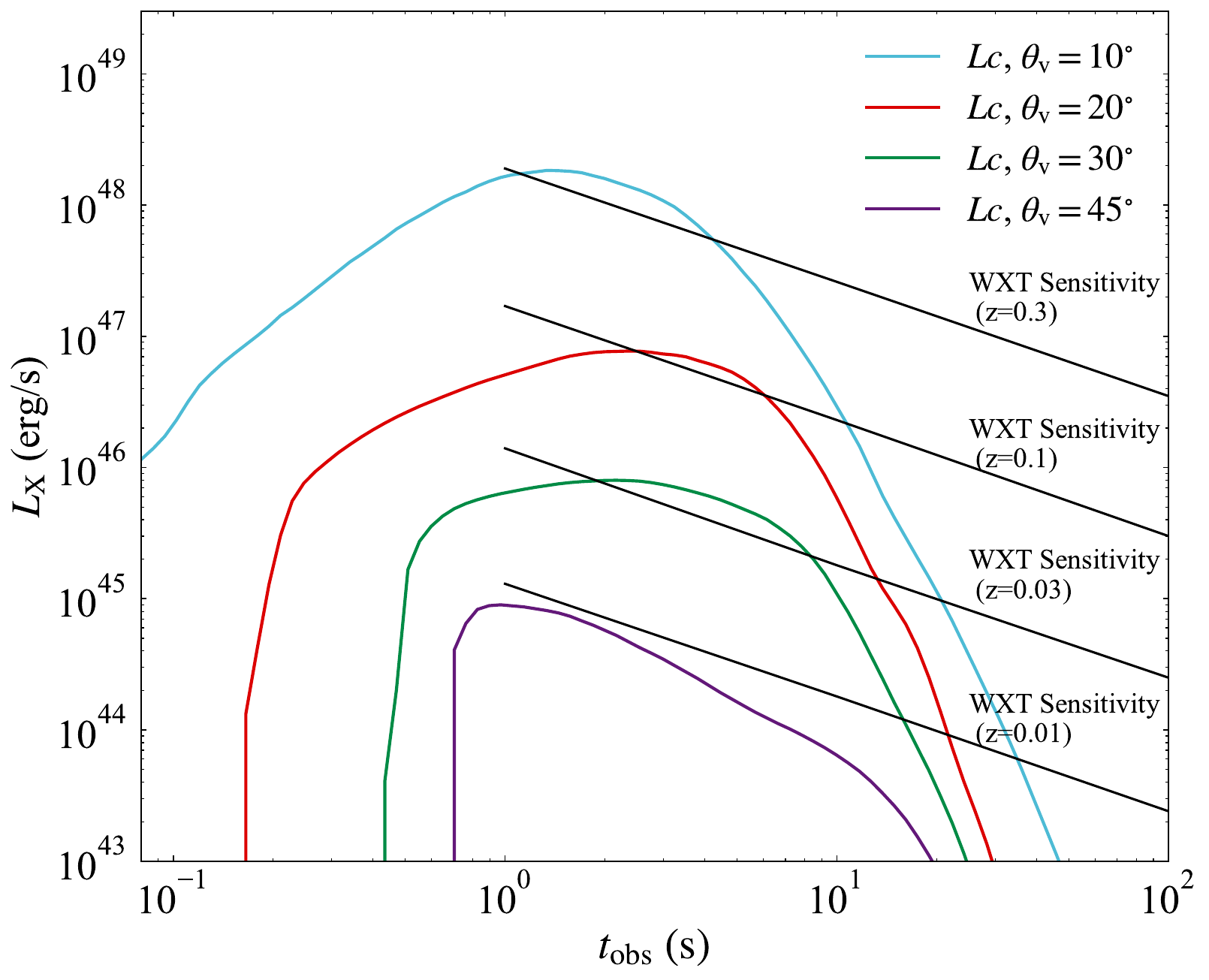}
    \includegraphics[width=0.32\textwidth]{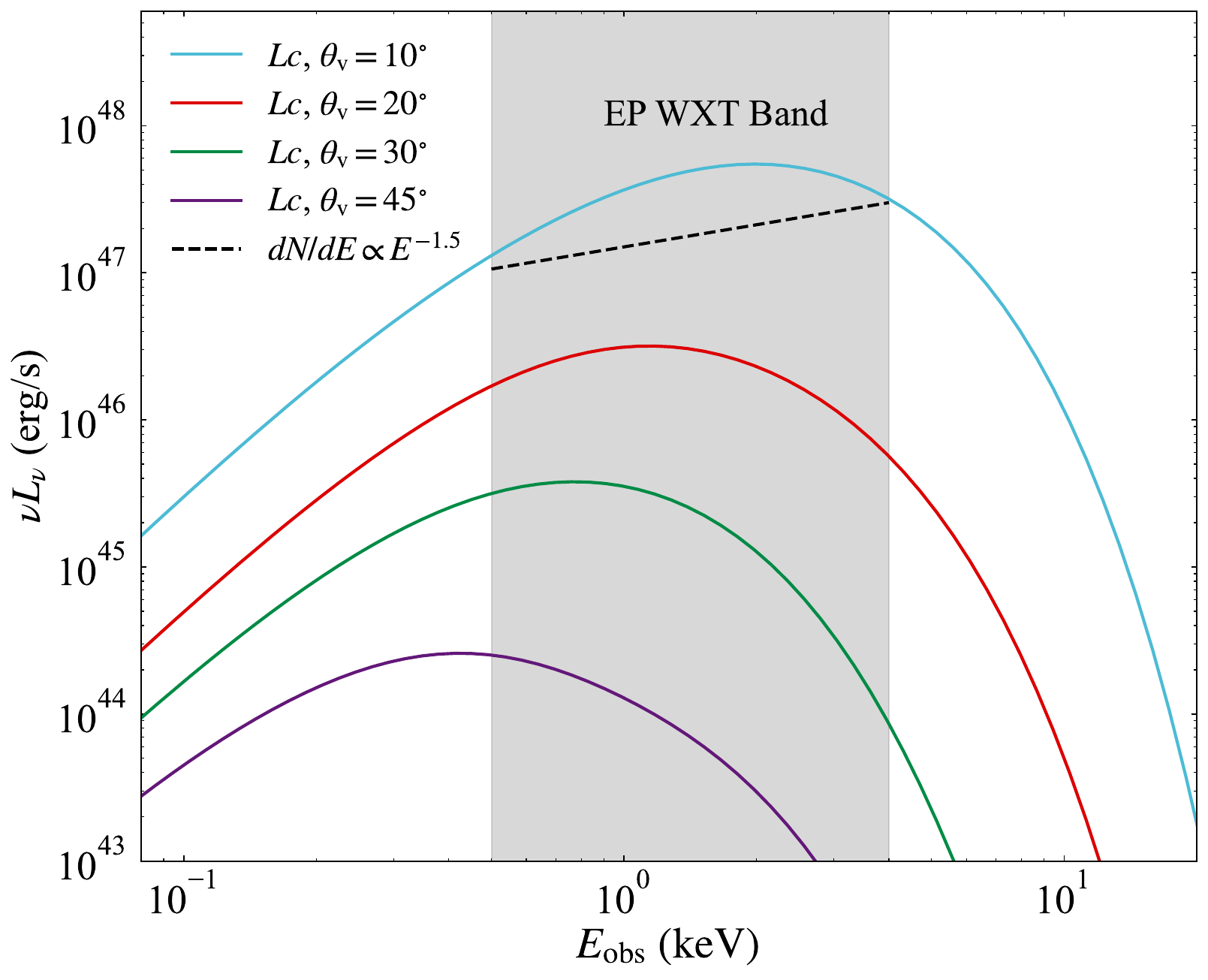}
    \includegraphics[width=0.32\textwidth]{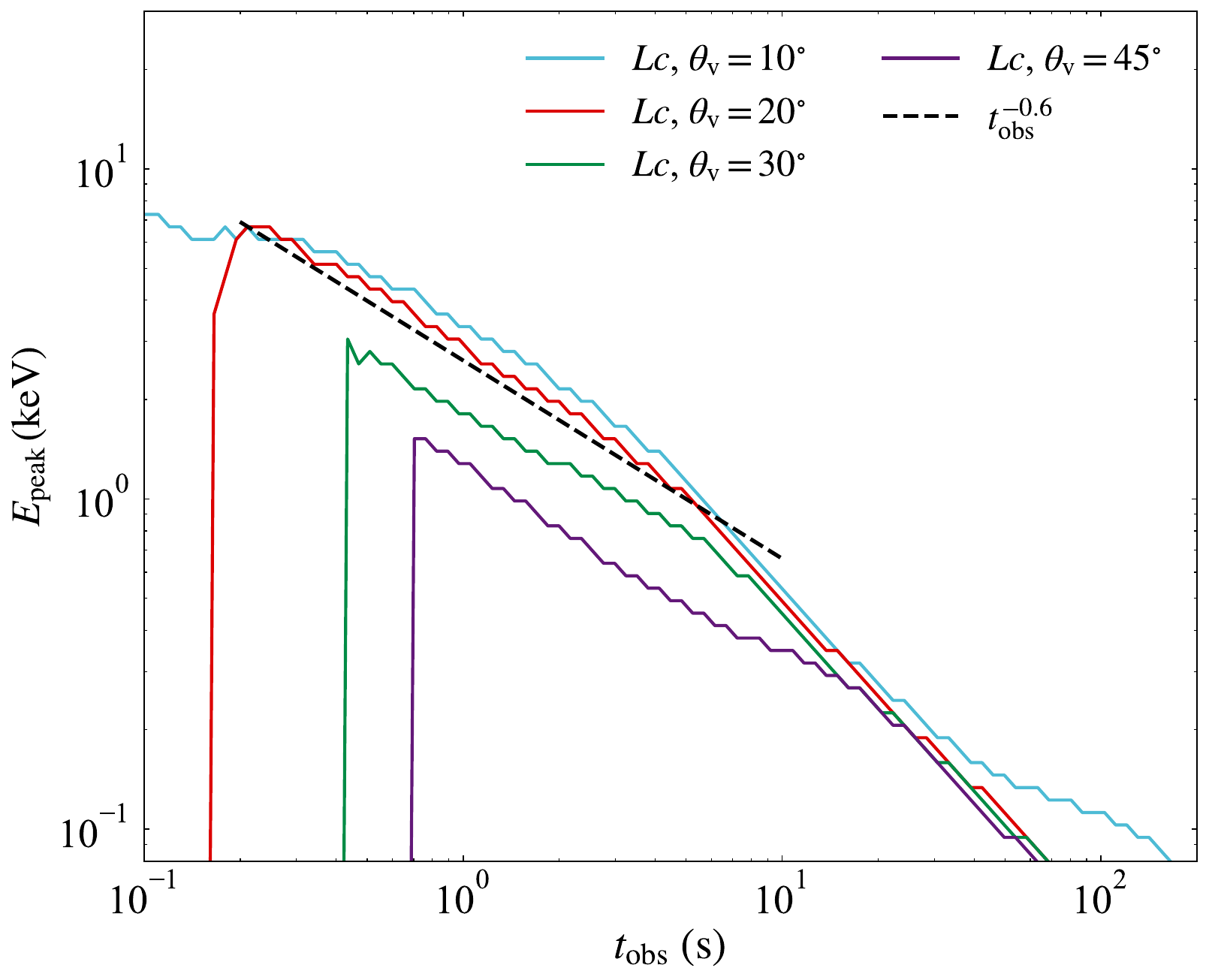}
    \caption{
    X-ray lightcurves and spectra of the canonical model ($Lc$). \textit{Left panel}: X-ray lightcurves in the EP-WXT band (0.5-4 keV). The blue, red, green, and purple lines correspond to the viewing angles of $10^{\circ}$, $20^{\circ}$, $30^{\circ}$, and $45^{\circ}$. The black solid lines indicate the EP-WXT sensitivity limits taken from \cite{Yuan2022hxga.book...86Y} for sources at redshifts $z=0.01$--0.3.
    \textit{Middle panel}: The $T_{90}$-averaged spectra at different viewing angles. The gray shaded region marks the EP WXT band, while the black dashed line denotes a power-law spectrum with a photon index of 1.5.
    \textit{Right panel}: The temporal evolution of the peak energy in the $Lc$ model.
    The black dashed line indicates a power-law decay $E_{\rm peak}\propto t^{-0.6}_{\rm obs}$.    
    }
    \label{fig:Lc}
\end{figure*}

We assume that the jet power remains constant for a duration of $t_{\rm j}\approx1\,$s in the lab frame. Then the injection rapidly cuts off after $t_{\rm j}$. At $t>t_{\rm j}$, we adopt an exponential cutoff for the jet velocity and pressure as in our previous work \citep{Zheng2026ApJ..1003L..19Z}.
We run the simulations up to a maximum lab-frame time of $t_{\rm max} = 5000\rm\,s$. By this stage, the radiation inside the cocoon has already diffused out.

After the hydrodynamic simulations, we calculate the cocoon cooling emission using the post-processing method developed in \citet{Zheng2026ApJ..1003L..19Z}.
{ To isolate the cooling component, we include only snapshots after the shock-breakout phase in the post-processing.}
At each lab-frame time, we compute the radial optical depth along each polar angle and determine the photospheric radius ($r_{\rm ph}$) and diffusion radius ($r_{\rm diff}$). The diffusive flux is estimated from the radiation energy density near $r_{\rm diff}$, and the escaping radiation is assumed to have a blackbody spectrum in the comoving frame (see Appendix \ref{app:LTE} for a discussion) with temperature determined by the local energy density. 
{ We verify in Appendix \ref{app:LTE} that the cocoon regions responsible for the cooling emission satisfy the local thermodynamic equilibrium (LTE) condition. This differs from the low-density shock-breakout layer, which can be photon-starved and produce a substantially harder spectrum \citep{Gottlieb2018CocoonSBO,gutierrez2025PhRvD.111f3031G}.}
We then transform the emission to the observer frame and integrate over the equal-arrival-time surface for a given viewing angle ($\theta_{\rm v}$). The zero point of the observer time is the breakout time.

In this Letter, we adopt the Rosseland-mean opacity $\kappa_{\rm R}=0.2\,{\rm cm^2\,g^{-1}}$, { corresponding to the electron-scattering opacity of hot, fully ionized, hydrogen-poor material}. At lower temperatures, recombination makes bound-bound opacity important, especially in NSM ejecta enriched by $r$-process elements \citep{Kasen2013ApJ...774...25K,Banerjee2026arXiv260120495B}. Thus, our early X-ray predictions are robust, while the early UV emission from SGRB cocoons may be substantially suppressed \citep{Banerjee2026arXiv260120495B}.

\section{Results}
\label{sec:res}

\subsection{Cocoon Physics}
\label{sec:phy}
Figure \ref{fig:Engden} illustrates the evolution of the jet-cocoon system in our canonical model $Lc$. The NSM ejecta are initially anisotropic, with a lower density along the polar direction (see the upper left panel in Figure \ref{fig:Engden}). The jet head propagates at a trans-relativistic velocity within the ejecta and breaks out from the ejecta at lab-frame time $t_{\rm bo}\approx0.3\,$s. After breakout, the cocoon expands laterally and progressively becomes quasi-spherical. Meanwhile, the jet continues to move outward, eventually forming a thin relativistic shell ahead of the system. The cocoon then expands nearly radially, while the slower NSM ejecta trail behind.

In all models considered in this work, the jet successfully breaks out of the NSM ejecta, and therefore the cocoon exhibits a broadly similar angular structure. Figure~\ref{fig:angular} shows the angular distributions of the isotropic-equivalent energy $E_{\rm iso}(\theta)=\int^{R_{\rm c,max}}_{v_{\rm max}t}(T_{00}-\Gamma\rho'c^2)4\pi r^2dr$, and the photospheric 4-velocity $u_{\rm ph}=\Gamma_{\rm ph}\beta_{\rm ph}$, when the cocoon front reaches $R_{\rm c,max}\approx10^{13}\,$cm. Here, $E_{\rm iso}$ is calculated by integrating the energy between the NSM ejecta and the cocoon front, excluding the contribution from the NSM ejecta itself.

Most models display a bright, nearly uniform core extending to $\theta\sim3^\circ$, followed by a sharp decline at $\theta\approx3^\circ-10^\circ$ and a slowly decaying cocoon component at larger angles ($\theta\approx10^\circ-40^\circ$).  
For most models, the cocoon at $\theta =10^\circ$ has an isotropic-equivalent energy of $E_{\rm iso}\approx10^{50}\,$erg and a photospheric Lorentz factor of $\Gamma_{\rm ph}\approx8$. Since the jet typically breaks out from the ejecta at a radius of $R_{\rm bo}\approx v_{\rm max}(t_{\rm delay}+t_{\rm bo})\sim10^{10}\,$cm, the peak luminosity and duration of the cocoon can be estimated by considering the cocoon expansion from $R_{\rm bo}$ up to the diffusion radius, and we obtain \citep{Zheng2025ApJ...985...21Z}
\begin{equation}
    \label{eq:lumi}
    \begin{split}
    L_{\rm peak} &\approx 5\times10^{47}\,R_{\rm bo,10}\Gamma_{\rm ph,1}^{13/3}\,{\rm erg\,s^{-1}} \\
   t_{\rm peak} &\approx (3\,{\rm s})E_{\rm iso,50}^{1/2}\Gamma_{\rm ph,1}^{-5/2}.
    \end{split}
\end{equation}
The main deviation from this common behavior is the powerful jet model ($Lp$), whose energy budget is larger, resulting in a higher $E_{\rm iso}$ and brighter cocoon emission (see below).

\subsection{X-ray Emission}
\subsubsection{General Properties}
We show the X-ray lightcurves and spectra observed at different viewing angles $\theta_{\rm v}$ for our canonical model $Lc$ in Figure \ref{fig:Lc}. The X-ray luminosity $L_{\rm X}(t_{\rm obs})$ at a given observer time $t_{\rm obs}$ is integrated over the EP WXT band (0.5–4 keV), {while ignoring cosmological redshift factors (which only weakly affect our results for the anticipated low-redshift sources $z\lesssim 0.3$)}.
We find that the WXT X-ray lightcurves rise rapidly within $\sim$0.3 seconds and last for 1-10 s, depending on the viewing angles. After the peak, the luminosity $L_{\rm X}$ declines rapidly because the emitting cocoon cools rapidly and the spectral peak shifts below the WXT band, which is supported by the evolution of the peak energy shown in the right panel of Figure \ref{fig:Lc}.

For a slightly off-axis observer at $\theta_{\rm v}=10^{\circ}$, the peak luminosity reaches $\sim 10^{48}\,{\rm erg\,s^{-1}}$ with a duration of $\sim\!10\,$s, making it detectable for EP WXT up to a redshift of $z\approx 0.3$. As the viewing angle increases, both the peak luminosity and duration decrease because there is less relativistic emitting gas. 
FXTs seen from large viewing angles ($\theta_{\rm v}\geq45^{\circ}$) are much fainter ($L_{\rm X}\lesssim10^{45}{\rm erg\,s^{-1}}$), and would only be marginally detectable for nearby events that are similar to GW170817 ($z\approx0.01$,$D_{\rm L}\approx40\,$Mpc).

\begin{figure}
    \centering
    \includegraphics[width=0.9\linewidth]{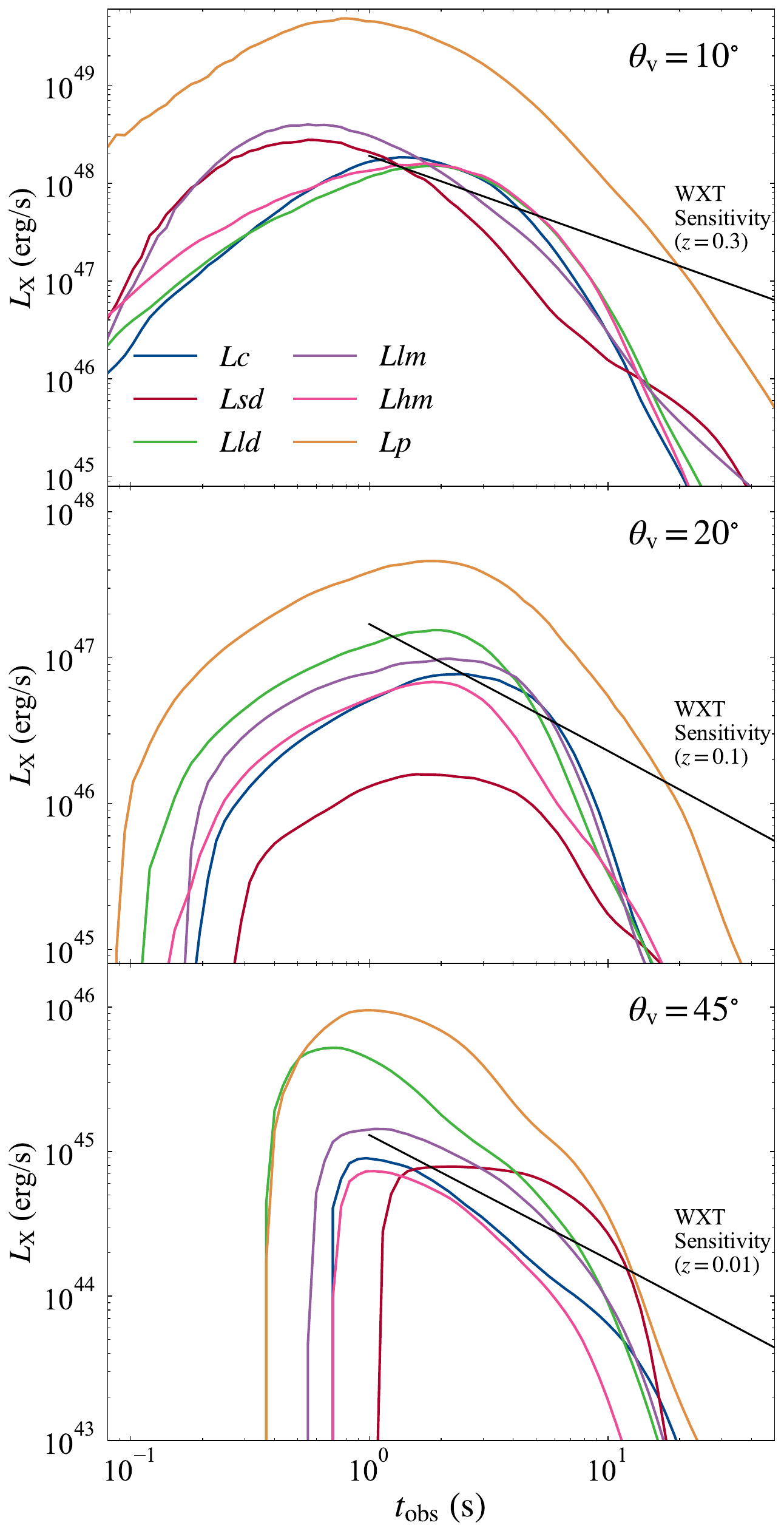}
    \caption{EP WXT (0.5-4 keV) lightcurves in different models. Colors represent different models, as in Figure \ref{fig:angular}.
    The upper, middle, and lower panels correspond to viewing angles of $\theta_{\rm v}=10^{\circ}$, $20^{\circ}$, and $45^{\circ}$, respectively.
    }
    \label{fig:LXcom}
\end{figure}

The middle panel of Figure~\ref{fig:Lc} shows the $T_{90}$-averaged spectra.  Here, $T_{90}$ is defined as the time interval between the epochs when 5\% and 95\% of the total energy is released in the WXT band. Similar to the collapsar cocoon models, the spectra become progressively softer as the viewing angle increases.
However, the NSM cocoon spectrum is harder (i.e., smaller photon index) in the WXT band than that of the collapsar cocoon at the same viewing angle. For example, at $\theta_{\rm v}=10^\circ$, the photon index of the $Lc$ model is $\sim$1.5, whereas the corresponding value for the collapsar cocoon is $\sim$2.5. 
This is because the NSM cocoon emission has a higher peak energy than that of the collapsar cocoon by about a factor of two (see Appendix~\ref{app:comparison} for details). 
Since the WXT band (0.5-4 keV) lies close to the peak of the quasi-thermal spectrum, even this modest shift in $E_{\rm peak}$ leads to a significant difference in the photon index. However, the cocoon emission is still considerably softer than the prompt jet emission. 
In particular, the hard prompt emission of GRB 170817A should not be directly compared with this cooling component, as it has been associated with jet or cocoon shock-breakout emission. Typical short GRBs have a low-energy photon index of $0.46^{+0.37}_{-0.68}$ in the Band function \citep{Poolakkil2021ApJ...913...60P}, while the short GRB detected by EP (GRB 250704B) has a comparable photon index of $0.78^{+0.27}_{-0.31}$ \citep{Li2026_250704B}.

Another important property of the X-ray spectrum is that $E_{\rm peak}$ decreases monotonically with time, approximately as $t^{-0.6}_{\rm obs}$, which is consistent with the behavior found in collapsar cocoon models. We conclude that NSM and collapsar cocoons show similar viewing-angle and temporal trends, but differ in their peak energy.

\begin{figure}
    \centering
    \includegraphics[width=\linewidth]{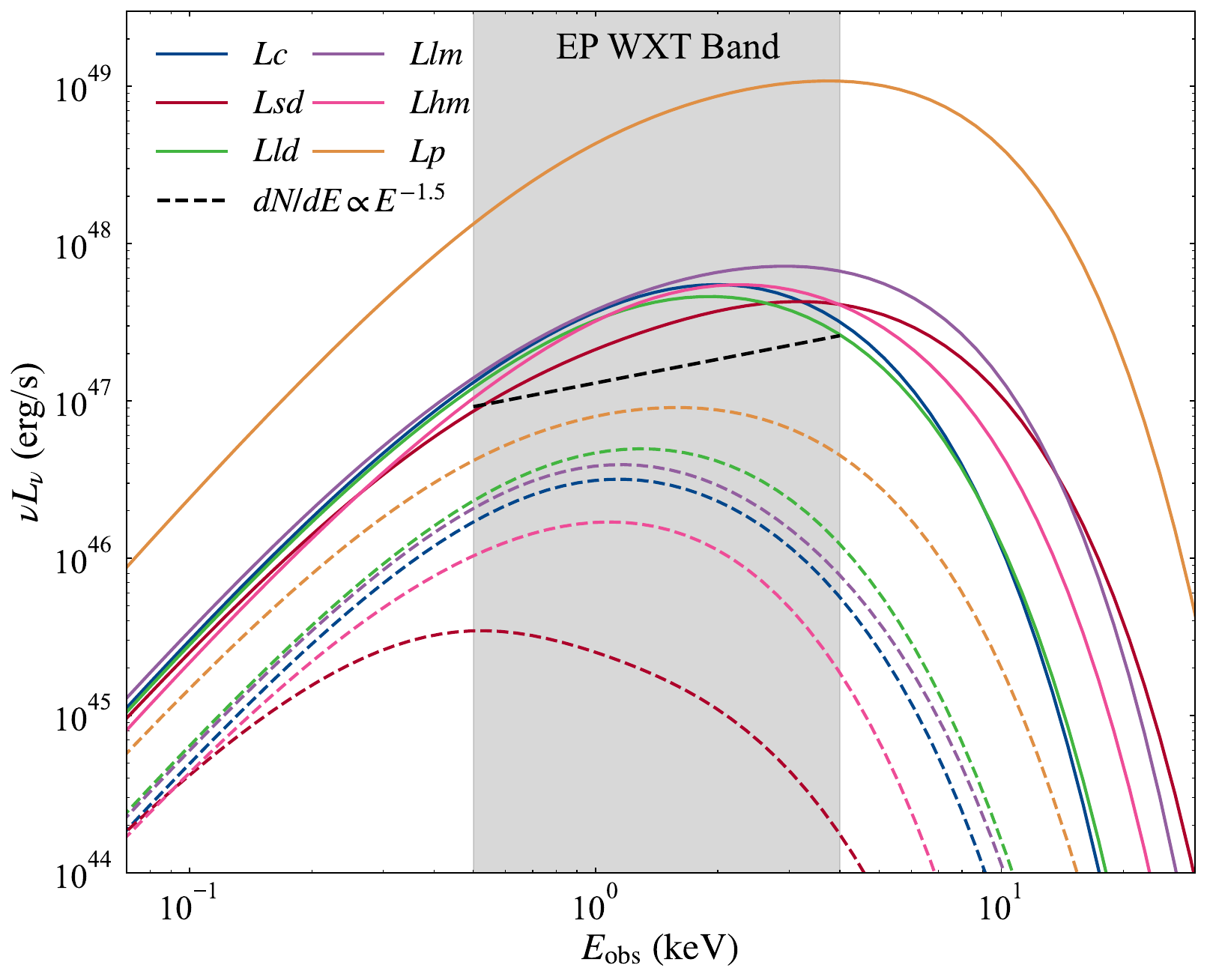}
    \caption{The $T_{90}$-averaged spectra for different models. Colors represent different models, as in Figure \ref{fig:angular}. The solid and dashed curves correspond to viewing angles of $\theta_{\rm v}=10^{\circ}$ and $20^{\circ}$, respectively. The black dashed line shows a power-law spectrum with a photon index of $1.5$.
    }
    \label{fig:avgmix}
\end{figure}

\subsubsection{Model Comparison}
Below we examine how the cocoon emission depends on the model parameters listed in Table \ref{tab:param}. Results are shown in Figures \ref{fig:LXcom} and \ref{fig:avgmix}. We investigate the effects of three parameters: the jet power, the ejecta mass, and the delay time between the merger and jet launching. Among these parameters, the jet power has the strongest impact on the cocoon emission with more powerful jets producing brighter cocoons. In comparison, a longer delay time leads to a moderate enhancement of the cocoon emission, while different ejecta mass has only a minor effect over the parameter space we explored in this Letter.

The impact of different parameters can be understood from the cocoon energy budget.
Before breakout, the jet continuously deposits its energy into the surrounding NSM ejecta, leading to a cocoon energy of $E_{\rm c}\simeq L_{\rm j}(t_{\rm bo}-R_{\rm bo}/c),$ where $L_{\rm j}$ is the jet power, $t_{\rm bo}$ is the breakout time, and $R_{\rm bo}$ is the breakout radius. 
Because the jet head already propagates at a trans-relativistic $v\sim 0.6c$ in the canonical model ($Lc$), a more powerful jet cannot reduce the breakout time substantially. Therefore, the breakout times of the $Lc$ model (0.3 s) and the $Lp$ model (0.2 s) are comparable (see Table \ref{tab:param}), and the cocoon energy is roughly proportional to the jet power, leading to a more energetic cocoon in the $Lp$ model. Consequently, the $Lp$ model has larger $E_{\rm iso}$ and $u_{\rm ph}$, producing brighter cocoon emission at all viewing angles in Figure \ref{fig:LXcom}.

The jet delay time $t_{\rm delay}$ directly affects the breakout radius via $R_{\rm bo}\approx v_{\rm max}(t_{\rm delay}+t_{\rm bo})$. A longer delay time results in a longer breakout time, leading to a more energetic cocoon. Considering the breakout time of the long delay ($Lld$) model (0.7 s) and the short delay ($Lsd$) model (0.2 s), we find the cocoon energy to be approximately twice as large in the $Lld$ case, leading to brighter emission at most viewing angles.

\begin{figure}
    \centering
    \includegraphics[width=\linewidth]{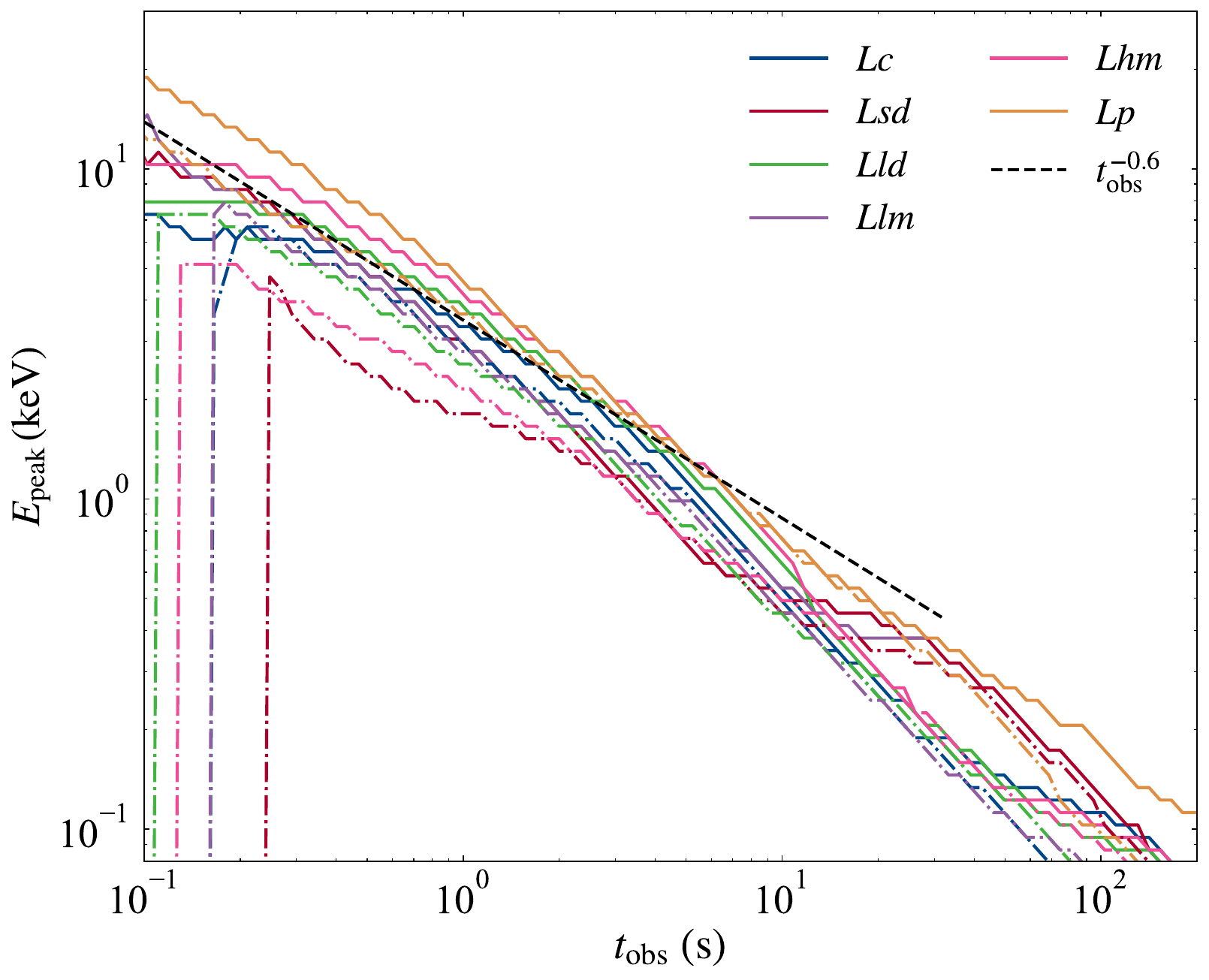}
    \caption{
    Temporal evolution of the peak energy for different models. Colors denote different models, as in Figure \ref{fig:angular}. 
    The solid and dash-dotted curves correspond to viewing angles of $\theta_{\rm v}=10^{\circ}$ and $20^{\circ}$, respectively. The black dashed line indicates the power-law decay $E_{\rm peak}\propto t_{\rm obs}^{-0.6}$.
    }
    \label{fig:Epeakmix}
\end{figure}

The impact of the ejecta mass is minor because the breakout time is comparable in the low mass ($Llm$) case (0.2 s) and high mass ($Lhm$) case (0.4 s), leading to comparable cocoon energy. The smaller ejecta mass results in a smaller cocoon mass, leading to a higher Lorentz factor while the cocoon energy is comparable. The angular distribution of $E_{\rm iso}$ and $u_{\rm ph}$ in Figure \ref{fig:angular} confirms that the $Llm$ case has larger $u_{\rm ph}$. Hence, the cocoon emission in the $Llm$ case is slightly brighter than that of $Lhm$ case because the luminosity is more sensitive to the Lorentz factor (see Eq. \ref{eq:lumi}).

Among all models, the main observational spectral features are qualitatively similar. The photon index of the $T_{90}$-averaged spectra is $\sim$1.5 across the WXT band for small viewing angle ($\theta_{\rm v}=10^{\circ}$) and increases (i.e., getting softer) when the viewing angle increases (see Figure \ref{fig:avgmix}).
The X-ray spectra soften over time during the $T_{90}$ interval, and the time evolution of the peak energy in all models is shown in Figure \ref{fig:Epeakmix}. Overall, $E_{\rm peak}$ decreases roughly as $\sim t^{-0.6}_{\rm obs}$ during the FXT phase ($t_{\rm obs}\lesssim10\,$s) in all models and viewing angles.

\begin{figure}
    \centering
    \includegraphics[width=\linewidth]{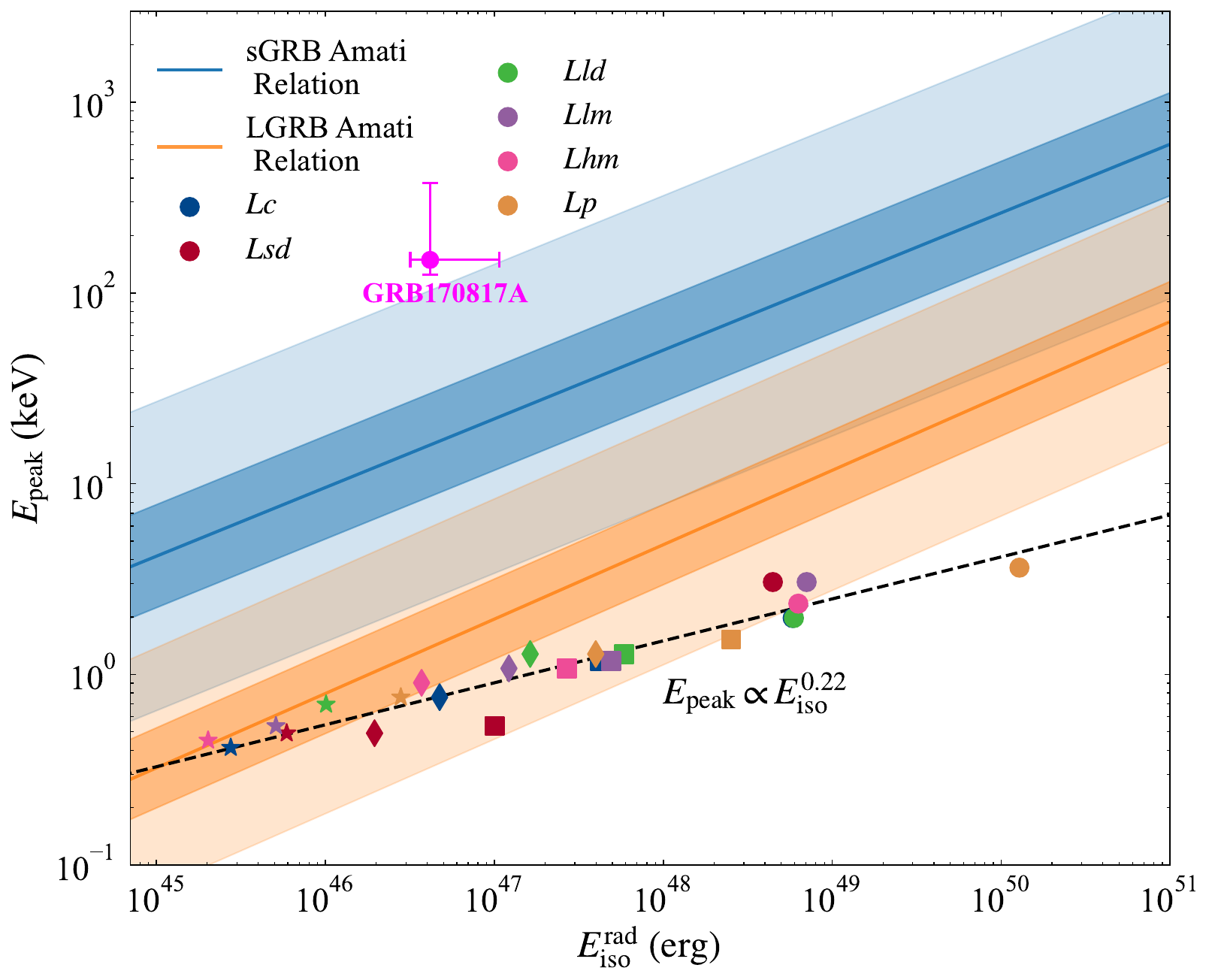}
    \caption{
    The $E^{\rm rad}_{\rm iso}$ vs. $E_{\rm peak}$ diagram. The blue solid and orange lines show the Amati relation for the SGRB (Type-I GRB) and LGRB (Type-II GRB) populations, adopted from \cite{Liu2025NatAs...9..564L}. The dark- and light- shaded regions indicate the $1\sigma$ and $3\sigma$ confidence intervals, respectively.  { The magenta point marks GRB 170817A.}
    Colors represent different models, while the circles ($\theta_{\rm v}=10^{\circ}$), squares ($\theta_{\rm v}=20^{\circ}$), diamonds ($\theta_{\rm v}=30^{\circ}$), and stars ($\theta_{\rm v}=45^{\circ}$) denote different viewing angles. 
    }
    \label{fig:Amati}
\end{figure}

In Figure \ref{fig:Amati}, we show the $E^{\rm rad}_{\rm iso}$ vs. $E_{\rm peak}$ correlation and compare it with the so-called Amati relation inferred from SGRB samples \citep{Liu2025NatAs...9..564L}. Here, $E_{\rm peak}$ is the peak energy of the $T_{90}$-averaged spectrum shown in Figure \ref{fig:avgmix} and $E^{\rm rad}_{\rm iso}$ is the isotropic equivalent total radiated energy obtained by integrating the WXT lightcurves $L_{\rm X}(t_{\rm obs})$. 
We find that FXTs from cocoon cooling emission are a distinct population compared to the non-thermal gamma-ray prompt emission, with a correlation 
\begin{equation}
    (E_{\rm peak}/1.5{\rm \, keV})\approx (E^{\rm rad}_{\rm iso}/10^{48}{\rm \, erg})^{0.22}.
\end{equation}
For comparison, the Amati relation for SGRBs is $(E_{\rm peak}/580{\rm \, keV})\approx (E^{\rm rad}_{\rm iso}/10^{51}{\rm \, erg})^{0.36}$ \citep{Liu2025NatAs...9..564L}. The weak $E_{\rm peak}$--$E^{\rm rad}_{\rm iso}$ correlation of NSM cocoon emission is slightly different from that for LGRB cocoons (where $E_{\rm peak}$ stays nearly constant). This difference arises from the anisotropic density profile of the NSM ejecta (see Appendix~\ref{app:comparison} for details).

\section{Luminosity Function and Detection Rate}
\label{sec:rate}
In this section, we calculate the EP detection rate of NSM cocoons. The simplest estimate for the detection rate is given by
\begin{equation}
    \dot{N}_{\rm det}\approx f_{\rm duty}f_{\rm sky}V_{\rm max}\dot{R}_{\rm cocoon},
\end{equation}
where $f_{\rm duty}=0.5$ is the duty cycle of EP, $f_{\rm sky}=0.087$ is the fraction of the sky covered by EP WXT \citep{Yuan2025SCPMA..6839501Y}, $V_{\rm max}\approx4\pi D_{\rm c}^3/3$ is the maximum volume within which the cocoon is detectable, and $\dot{R}_{\rm cocoon}$ is the volumetric NSM cocoon event rate with favorable viewing angle for detection.

Because the brightest cocoon emission at small viewing angles $\theta_{\rm v}\simeq 10^\circ$ has the largest detectable volume, it dominates the detection rate. Assuming that all NSMs launch successful jets, the event rate density of the cocoon at $\theta_{\rm v}=10^{\circ}$ is $\dot{R}_{\rm cocoon}=f_{\rm b}\dot{R}_{\rm NSM}$, where $f_{\rm b} = \theta_{\rm v}^2/2\approx0.015$ is the beaming factor (accounting for both jets) and $\dot{R}_{\rm NSM}$ is the volumetric neutron star merger rate. The latest GW survey has constrained the NSM rate to be $\dot{R}_{\rm NSM}=59.3^{+95.4}_{-43.9}\,{\rm Gpc}^{-3}{\rm yr}^{-1}$ \citep{LIGO2026GWTC_5}. Therefore, the cocoon event rate is of the order $\dot{R}_{\rm cocoon}\sim 1\,{\rm Gpc}^{-3}{\rm yr}^{-1}$. Considering the brightest cocoon is detectable at $z\approx0.3$, corresponding to $V_{\rm max}\approx7\,{\rm Gpc}^{3}$, we therefore obtain a detection rate of $\dot{N}_{\rm det}\sim 0.3\,{\rm yr}^{-1}$ for EP WXT.

Then we perform a more detailed numerical calculation for the detection rate by including all viewing angles spanning from $10^{\circ}$ to $55^{\circ}$. The contribution from larger viewing angles $>55^\circ$ is negligible.
We divide the angular region into independent rings. Each ring has its own beaming factor and maximum detectable redshift because the observed luminosity decreases with increasing viewing angle. 
The full-sky event rate of the $i$-th ring is
\begin{equation}
    {{\dot N}_{\rm ring,i}} = {f_{\rm b,i}}{{\dot R}_{\rm NSM}}\int_0^{{z_{\rm \max ,i}}} {\frac{1}{{1 + z'}}\frac{{4\pi cD_c^2\left( {z'} \right)}}{{H\left( {z'} \right)}}} dz',
\end{equation}
where $f_{\rm b,i}$ is the beaming factor of the ring and the factor $(1+z)^{-1}$ accounts for cosmological time dilation. The integral is the full expression for the comoving volume. Here we neglect the redshift evolution of the NSM rate because the cocoon is detectable only in the nearby universe where the evolution is expected to be weak. 
For each ring, we compute the X-ray lightcurves at the viewing angle of $\theta_{\rm v}=(\theta_{\rm i}+\theta_{\rm i+1})/2$. Then we compare the lightcurves with the WXT sensitivity to determine the maximum detectable redshift $z_{\rm max,i}$ for the ring.

The total cocoon detection rate is obtained by summing the contributions from all angular rings,
\begin{equation}
    {{\dot N}_{\rm \det }} = {f_{\rm sky}}{f_{\rm duty}}\sum\limits_i {{{\dot N}_{\rm ring,i}}}.
\end{equation}

For our canonical model ($Lc$), the predicted EP WXT detection rate is 
\begin{equation}
    {{\dot N}_{\rm \det }}\approx 0.5 \left(\frac{{\dot R}_{\rm NSM}}{100\,{\rm Gpc}^{-3}{\rm yr}^{-1}} \right) {\rm yr}^{-1}.
\end{equation}
This result is consistent with our analytical estimate above. For our canonical model, we find that EP can detect only one NSM cocoon every $\sim 2$ years (depending on $\dot{R}_{\rm NSM}$). Since EP began operation in spring 2024, a first candidate might already be present in the archival data.
Because the predicted cocoon emission is faint and short-lived, some events may remain below the onboard trigger threshold. A systematic archival search for sub-threshold events provides an opportunity to discover the first NSM cocoon candidate.
The other models have comparable rates, except for $L_p$, whose higher luminosity increases the rate by an order of magnitude.

We further calculate the cumulative luminosity function of NSM cocoon emission. The local cumulative luminosity function is defined as the event rate of cocoons with peak observed luminosities above 
$L_{\rm peak}$ in the EP-WXT band,
\begin{equation}
    \Phi(>L_{\rm peak})=\int^{\infty}_{L_{\rm peak}}\frac{d \dot{R}_{\rm cocoon}}{dL'_{\rm peak}}dL'_{\rm peak},
\end{equation}
where $L_{\rm peak}$ is the peak luminosity in the WXT band and $\dot{R}_{\rm cocoon}$ is the local NSM cocoon event rate. $\Phi(>L_{\rm peak})$ gives the local volumetric rate of cocoons brighter than a given peak luminosity and describes how the event rate is distributed across luminosity thresholds.

\begin{figure}
    \centering
    \includegraphics[width=\linewidth]{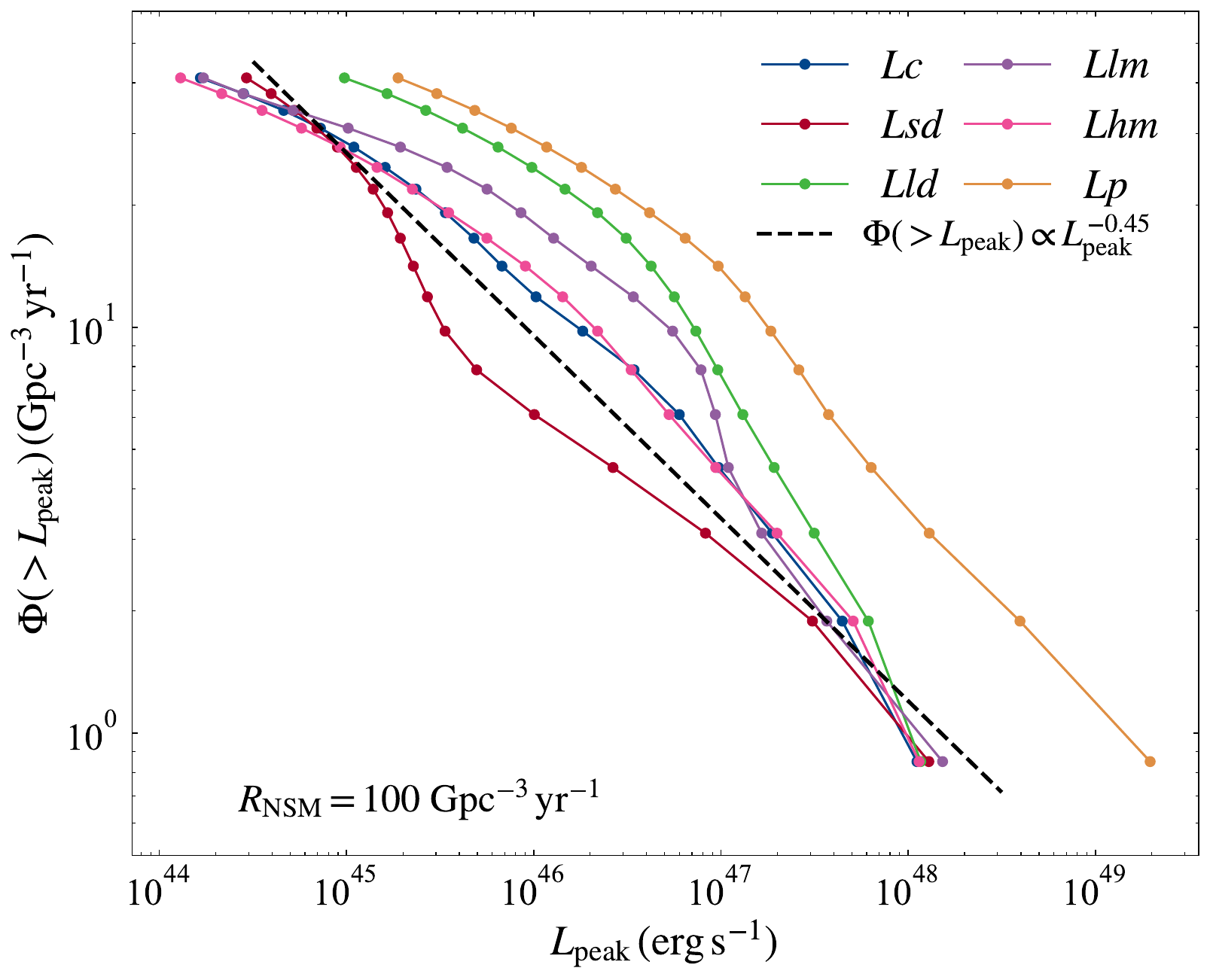}
    \caption{The cumulative luminosity function for different models. The solid curves show the luminosity functions as a function of the peak luminosity in the EP-WXT band, with colors denoting different models. The black dashed line indicates the trend $\Phi(>L_{\rm peak})\propto L_{\rm peak}^{-0.45}$. 
    }
    \label{fig:lumifun}
\end{figure}

In each of our models, the diversity in $L_{\rm peak}$ arises from the dependence on the viewing angle. The differential luminosity function can therefore be expressed as
\begin{equation}
\frac{d\dot{R}_{\rm cocoon}}{dL_{\rm peak}}
=\frac{d\dot{R}_{\rm cocoon}}{d\mu} \left| \frac{d\mu}{dL_{\rm peak}}\right|,
\end{equation}
where $\mu=\cos\theta_{\rm v}$. The relation between $L_{\rm peak}$ and $\theta_{\rm v}$ is obtained from the simulated lightcurves. Assuming that the jet axes of NSMs are isotropically oriented, the differential rate is constant $d\dot{R}/d\mu=\dot{R}/4\pi$, and the shape of the luminosity function is therefore determined by the angular dependence of $L_{\rm peak}(\theta_{\rm v})$. 


The cumulative luminosity functions of cocoons are shown in Figure \ref{fig:lumifun}. Their general behavior can be approximately described by a power-law function $\Phi(>L_{\rm peak})\propto L_{\rm peak}^{-0.45}$. 
The detailed shape of the luminosity function depends on the model parameters. In particular, the $Lsd$ model is systematically fainter at small viewing angles. In contrast, the $Lp$ model produces systematically brighter cocoon emission over a broad range of viewing angles, shifting its luminosity function toward the high-luminosity side.
{ We emphasize that these luminosity functions represent only the limited parameter space explored here. A broader distribution of parameters would likely broaden the predicted distribution.}

\section{Verification Strategy}
\label{sec:veri}

Because EP detects many X-ray transients, identifying the physical origin of a candidate NSM cocoon requires additional counterparts. { This is particularly important for events without a GW detection, since the FXT alone may not distinguish a successful off-axis jet from alternative models (e.g., choked jet).} Therefore, We focus on two verification pathways: joint GW--X-ray detection and follow-up searches for an associated kilonova and delayed jet afterglow. { The kilonova can establish the NSM origin, while the non-thermal afterglow provides an important consistency check of the successful off-axis jet interpretation.}

The most direct verification channel is a nearby NSM with a GW trigger, which requires a serendipitous EP-WXT detection of the FXT \citep[e.g.,][]{Chopra2026arXiv260716420C}. In this case, the GW signal can identify the merger origin and constrain the viewing angle geometry, while the EP-WXT detection provides an arcminute-scale X-ray localization for rapid electromagnetic follow-up. Prompt optical imaging from wide-field telescopes (e.g., LSST) is useful for identifying the optical counterpart, refining the localization to sub-arcsecond precision, and enabling precise host-galaxy association. 
{ Once an arcminute-scale EP-WXT localization is available, rapid target-of-opportunity observations with large-aperture telescopes can reach greater depth.}
However, the chance of simultaneous LIGO-EP coverage is roughly given by $f_{\rm duty} f_{\rm sky}\sim 4\%$. Furthermore, the cocoon emission may only be detectable by EP-WXT for sufficiently small viewing angles $\theta_{\rm v}\lesssim 30^\circ$--$40^\circ$. These orientations account for about half of the LIGO-selected events \citep{2011CQGra..28l5023S}, so the probability of co-detection is $\sim 2\%$. 
{ To reach a $\sim1\sigma$ ($\sim68\%$) probability of at least one co-detection, a sample of $\sim57$ LIGO-detected NSMs would be required.}

\begin{figure*}
    \centering
    \includegraphics[width=0.9\linewidth]{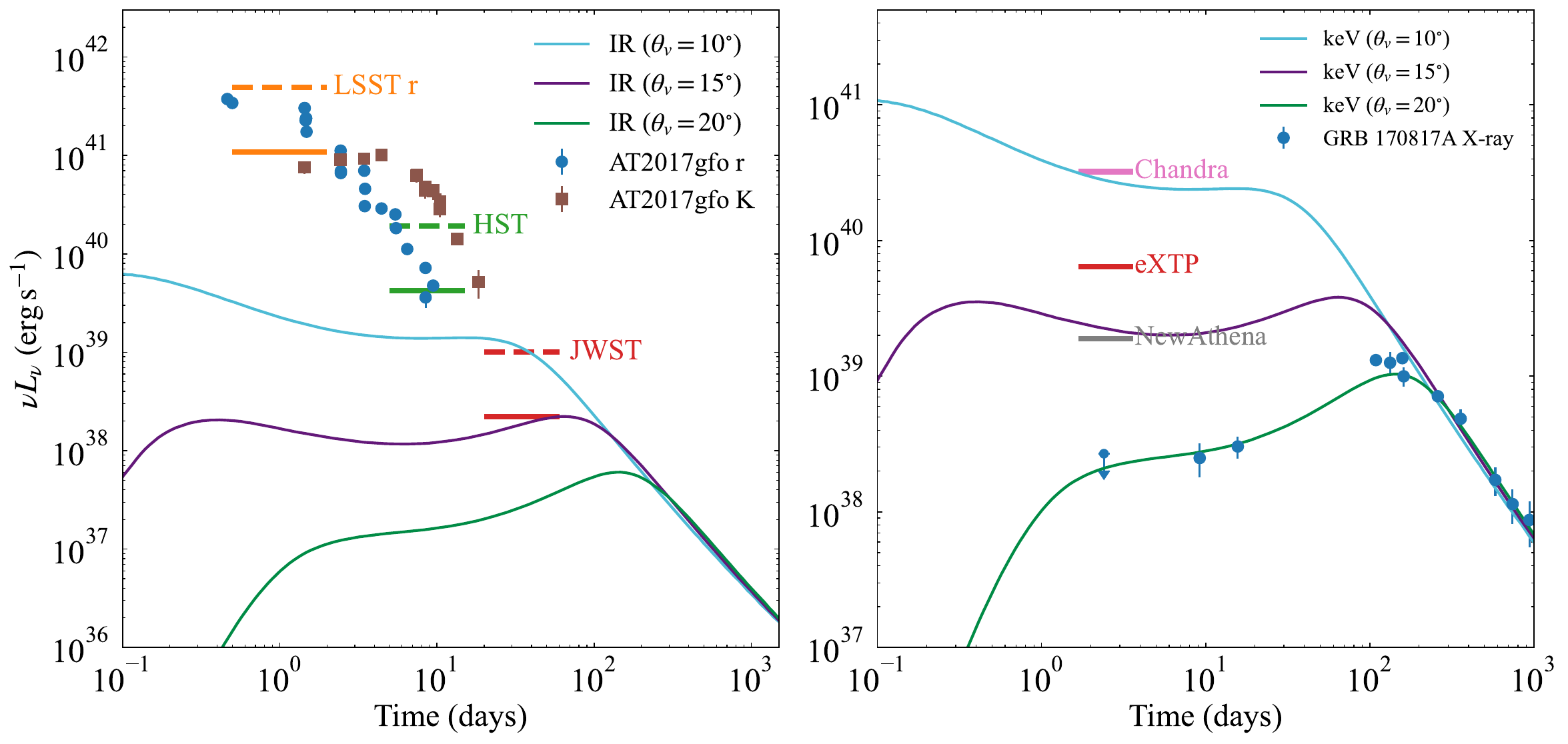}
    \caption{Detectability of the kilonova and afterglow. \textit{{ Left panel:}} The infrared afterglows at viewing angles $\theta_{\rm v}=10^{\circ}$, $15^{\circ}$, and $20^{\circ}$, together with the kilonova emission. The deep blue circles and brown squares show the $r$- and $K$-band observations of AT2017gfo, respectively. The { orange}, green, and red horizontal lines indicate the sensitivities of LSST $r$, HST F160W \citep{2025HST_WFC3IHB}, and JWST F150W2, respectively, at $z=0.1$ (solid lines) and $z=0.2$ (dashed lines). \textit{{ Right panel:}} The keV X-ray afterglow of the jet-cocoon at the same viewing angles. The deep blue circles show the X-ray observations of GRB 170817A. { The magenta, red, and gray horizontal lines show the approximate Chandra, eXTP, and NewAthena sensitivities, respectively, for a source at $z=0.1$. }}
    \label{fig:detect}
\end{figure*}


For events without a GW detection, an EP-triggered FXT can still be identified through its associated kilonova. Rapid { optical/infrared} follow-up is particularly useful for nearby candidates. An AT2017gfo-like kilonova at $z\lesssim0.1$ should be detectable by LSST in the $r$ band within the first day\footnote{See sensitivity at \href{https://rubinobservatory.org/for-scientists/rubin-101/key-numbers}{https://rubinobservatory.org/for-scientists/rubin-101/key-numbers}} (see the { left panel} of Figure \ref{fig:detect}). { Detection of such a counterpart would provide a precise localization, host-galaxy association, and evidence for an NSM origin.}

For comparison, the AT2017gfo $r$- and $K$-band photometry is shown in Figure \ref{fig:detect} \citep{diaz2017ApJ...848L..29D,Arcavi2017Natur.551...64A,Andreoni2017PASA...34...69A,Smartt2017Natur.551...75S,Tanvir2017ApJ...848L..27T,Cowperthwaite2017ApJ...848L..17C,Coulter2017Sci...358.1556C,Shappee2017Sci...358.1574S,Pian2017Natur.551...67P,Villar2017ApJ...851L..21V,Kasliwal2017Sci...358.1559K}.

At later times, deeper near-infrared observations with Hubble Space Telescope (HST) or James Webb Space Telescope (JWST) can search for the red kilonova component on timescales of several days to weeks\footnote{Sensitivity taken from \href{https://jwst-docs.stsci.edu/jwst-near-infrared-camera/nircam-performance/nircam-sensitivity}{https://jwst-docs.stsci.edu/jwst-near-infrared-camera/nircam-performance/nircam-sensitivity}.} (see Figure \ref{fig:detect}).

This phase is particularly important because several SGRBs, including GRB 050709 \citep[$z=0.16$,][]{Jin2016NatCo...712898J}, GRB 160821B \citep[$z=0.16$,][]{Lamb2019ApJ...883...48L,Troja2019MNRAS160821B}, and GRB 130603B \citep[$z=0.35$,][]{Tanvir2013Natur.500..547T} have shown kilonova-like optical/near-infrared excesses at comparable or greater distances. More recently, GRBs 211211A ($z=0.076$) and 230307A ($z=0.065$) provide additional nearby examples for identifying NSMs \citep{Yang2022Natur.612..232Y,Rastinejad2022Natur.612..223R,Levan2024Natur.626..737L,Yang2024Natur.626..742Y}. Thus, a rapidly evolving optical transient followed by a red excess would provide strong evidence that an EP FXT originates from an NSM.

{ At later times, the key test of the successful off-axis jet interpretation is the emergence of a delayed non-thermal afterglow.} In our model, the optical/infrared emission is dominated by the kilonova (or cocoon emission) during the first 10--20 d, while the afterglow becomes dominant after $\sim20$ days for $\theta_{\rm v}=10^\circ$--$20^\circ$ (Figure \ref{fig:detect}). { X-ray and radio follow-up on timescales of weeks to months can therefore provide a consistency check complementary to the early cocoon FXT. Thus, the early cocoon-cooling FXT and the delayed non-thermal afterglow are complementary signatures of the same successful structured jet.}
The GRB 170817A X-ray afterglow data shown in Figure 9 are from \citep{Margutti2017ApJ...848L..20M,Troja2017Natur.551...71T,Haggard2017ApJ...848L..25H,Margutti2018ApJ...856L..18M,Troja2019MNRAS170817,Nynka2018ApJ...862L..19N,Hajela2019ApJ...886L..17H,Makhathini2021ApJ...922..154M}.

We adopt the afterglow model from \cite{Zheng2025ApJ...985...21Z}, which successfully explains the GRB 221009A \citep{Zheng2024ApJ...966..141Z}. The angular profiles of the isotropic-equivalent kinetic energy $E_{\rm iso}(\theta)$ and initial four-velocity $u_{\rm 0}(\theta)$ are described by a power-law structured jet, given by
\begin{equation}
E_{\rm iso}(\theta)=
\frac{E_{\rm iso,max}}{\left[1+(\theta/\theta_{\rm c})^2\right]^{s_E/2}},
\,
u_0(\theta)=
\frac{u_{\rm 0,max}}{\left[1+(\theta/\theta_{\rm c})^2\right]^{s_u/2}},
\end{equation}
where $E_{\rm iso,max}$ and $u_{\rm 0,max}$ are the isotropic-equivalent energy and initial four-velocity at the jet core, $\theta_{\rm c}$ is the jet core angle, and $s_E=4.5$ and $s_u=2$ describe how rapidly the energy and four-velocity decrease outside the core. 
{ The adopted angular structure is consistent with our numerical simulations (see dashed lines in Figure \ref{fig:angular}) and is broadly consistent with the structured-jet interpretation of GRB 170817A \citep{Ryan2020ApJ...896..166R,Mooley2018Natur.561..355M,mooley2022_GW170817}.} Previous modeling and VLBI observations favor a core angle of a few degrees and a viewing angle of $\theta_{\rm v}\approx20^\circ$ for GRB 170817A \citep{Mooley2018Natur.561..355M,mooley2022_GW170817}, and our X-ray afterglow at this viewing angle is consistent with the observations.
In the fiducial model shown in Figure \ref{fig:detect}, we take $E_{\rm iso,max}=6\times10^{52}\,{\rm erg}$, $u_{\rm 0,max}=300$, $\theta_{\rm c}=3^{\circ}.5$, together with the ISM density $n_{\rm ISM}=10^{-3}\,{\rm cm}^{-3}$, electron index $p=2.2$, and microphysical parameters $\epsilon_e=0.1$ and $\epsilon_B=10^{-4}$.

For the GW170817-like parameters shown in Figure \ref{fig:detect}, the X-ray afterglow at $\theta_{\rm v}\approx10^\circ$ is only marginally detectable with Chandra at a few days (see the { right panel} of Figure \ref{fig:detect}). { The approximate effective areas of Chandra, eXTP, and NewAthena are $600$, $2800$, and $\sim10{,}000\,{\rm cm^2}$, respectively\footnote{See the effective area at \href{https://cxc.cfa.harvard.edu/cdo/about_chandra/}{Chandra X-ray Center instrument specifications}, \cite{Zhang2025eXTP..6819502Z}, and the \href{https://www.cosmos.esa.int/web/athena/about-athena}{ESA NewAthena mission page}.}, corresponding to ratios of roughly $1:5:17$. 
We estimate their relative sensitivities by scaling the Chandra limit according to these effective-area ratios. At $z=0.1$, NewAthena could also reach the fainter $\theta_{\rm v}=15^\circ$ afterglow around its peak. The $\theta_{\rm v}=20^\circ$ model remains below the approximate limits shown here.} The expected radio afterglow at 3 GHz is similarly faint, $F_{\nu}\approx10\,{\rm \mu Jy}$, below the three-hour sensitivity of the VLA.
{ Although a smaller viewing angle makes the afterglow brighter than that of GW170817, an event at $z\simeq0.1$ ($D_L\simeq460$ Mpc) is much more distant than GW170817 ($D_L\simeq40$ Mpc), reducing the observed flux by a factor of $\sim120$. Thus, the increase in afterglow brightness at smaller viewing angles does not necessarily compensate for the larger distance.}
Deep JWST observations can verify the afterglow component. For GW170817-like parameters, the infrared afterglow at $\theta_{\rm v}= 10^\circ$ and $15^\circ$ is detectable with JWST up to redshift $z\simeq0.2$ and $z\simeq0.1$, respectively.

We emphasize that Figure \ref{fig:detect} displays a GW170817-like model { as an illustrative benchmark}. The afterglow brightness depends sensitively on the jet angular structure, circumburst density, and microphysical parameters. Therefore, brighter X-ray or radio counterparts are possible for other NSM systems.

\section{Summary}
\label{sec:sum}
In this Letter, we perform numerical simulations of the jet-cocoon system in NSMs, following its evolution from jet launching to the cocoon diffusion radius ($\sim10^{13}\,\rm cm$). 
Using the post-processing method developed in our previous work, we calculate the cocoon cooling emission for different viewing angles and investigate its detectability and expected event rate with the Einstein Probe (EP) Wide-field X-ray Telescope (WXT).

Our calculations show that, for viewing angles of $\theta_{\rm v}=10^{\circ}$--$20^{\circ}$, the off-axis cocoon cooling emission can produce bright FXTs with luminosities of $L_{\rm X}\simeq10^{46-48}\,{\rm erg\,s^{-1}}$ and durations of $t_{\rm X}\simeq1$--$10\,{\rm s}$ in the WXT band. Such events are detectable by EP up to $z\approx 0.3$, corresponding to a detection rate of $0.5\,{\rm yr}^{-1}$, for an NSM rate of $\dot{R}_{\rm NSM}= 100\rm\, Gpc^{-3}\,yr^{-1}$. The observed spectra are quasi-thermal with a peak energy $E_{\rm peak}\simeq0.3$--$3\,{\rm keV}$. As the viewing angle increases, both the X-ray luminosity and emission duration decrease, and the peak energy also drops. At $\theta_{\rm v}=45^{\circ}$, the peak X-ray luminosity drops to $\sim\!10^{45}{\rm erg\,s^{-1}}$ and the duration shortens to $\sim3\,$s, limiting EP detection to nearby GW170817-like events ($z\lesssim0.01$).
{ At smaller viewing angles ($\theta_v\approx5^\circ -\!10^\circ$), the detector probes the jet–cocoon transition, where weak gamma-ray prompt emission from the jet and its high-latitude X-ray tail may coexist with the soft X-ray cocoon cooling. }

{ Much longer emission, such as the $560\,$s extended emission of GRB 250704B detected by EP \citep{Li2026_250704B}, would require a long-lasting central engine, such as a post-merger magnetar \citep{Zhang2013ApJ...763L..22Z} or an accreting black hole \citep{Rosswog2007MNRAS.376L..48R}.}

The X-ray spectra of cocoon cooling emission are systematically softer than the non-thermal ``prompt emission'' from GRB jets, and they soften with time. 
The $T_{90}$-averaged spectra have photon indices of approximately $1.5$ for $\theta_{\rm v}=10^{\circ}$, with larger (i.e. spectrally softer) photon indices at larger viewing angles.
Meanwhile, the peak energy decays as a power-law $E_{\rm peak}\propto t_{\rm obs}^{-0.6}$. These spectral properties are broadly similar to the jet-cocoon emission from collapsars. The main difference is that NSM cocoons have higher temperatures in $T_{90}$-averaged spectra, resulting in harder spectra in EP WXT band.

In this Letter, we focus primarily on the cooling X-ray from the cocoon, which is much less sensitive than the UV/optical emission to uncertainties in the wavelength-dependent opacity of the NSM ejecta. Our fiducial model overpredicts the early UV emission because it underestimates the UV opacity \citep{Banerjee2026arXiv260120495B}. 
A more realistic treatment of the ejecta composition and opacity is required to model the early UV/optical emission accurately and is left for future work.



\begin{acknowledgments}
We thank Bing Zhang and Xiang-Yu Wang for helpful discussions. J.-H.Z.’s work is supported by the National Natural Science Foundation of China (grant No. 124B2057).

\end{acknowledgments}

\appendix
\renewcommand{\thefigure}{A\arabic{figure}}
\setcounter{figure}{0}
\section{Comparison between LGRB and SGRB Cocoons}
\label{app:comparison}

\begin{figure}
    \centering
    \includegraphics[width=0.45\linewidth]{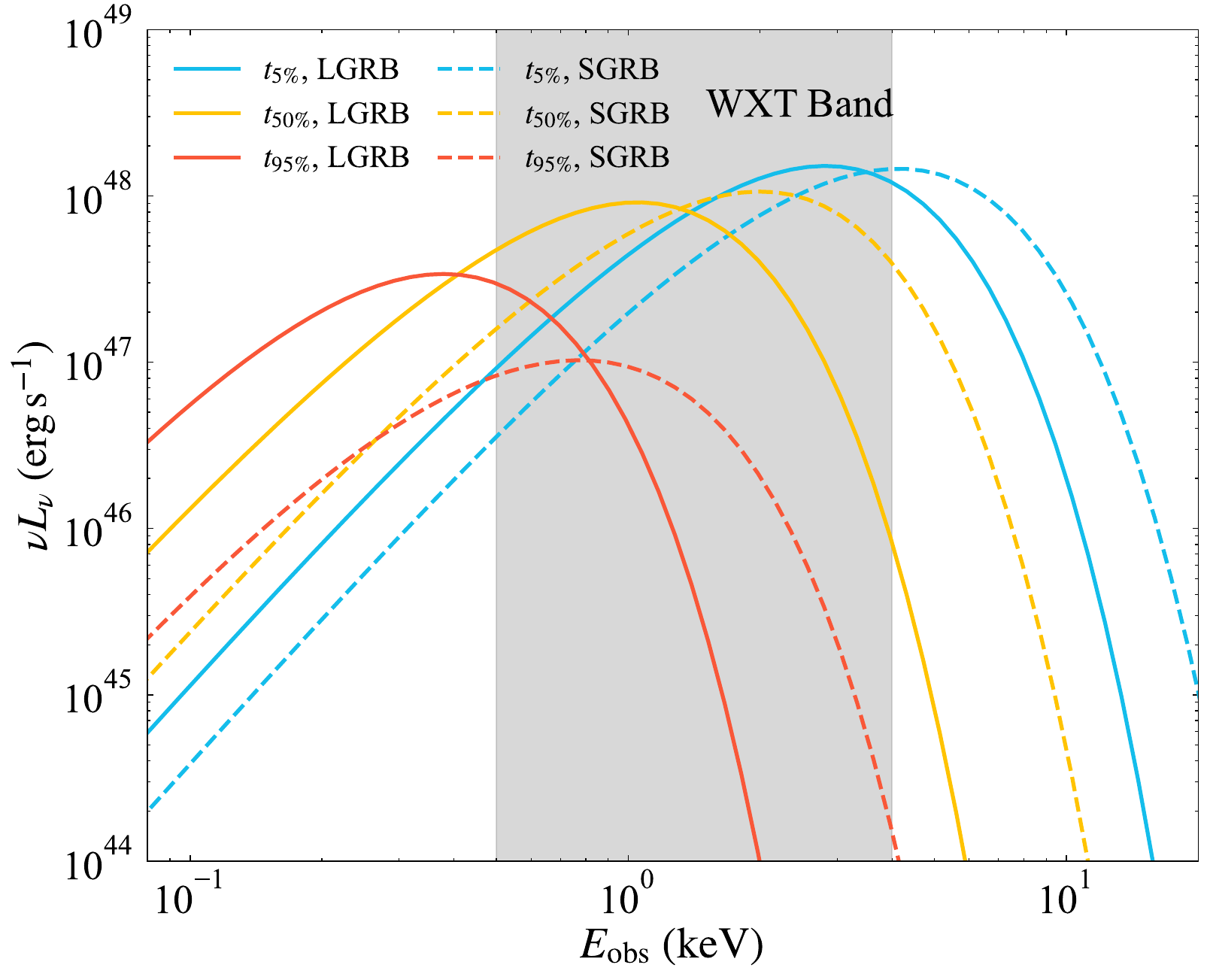}
    \includegraphics[width=0.45\linewidth]{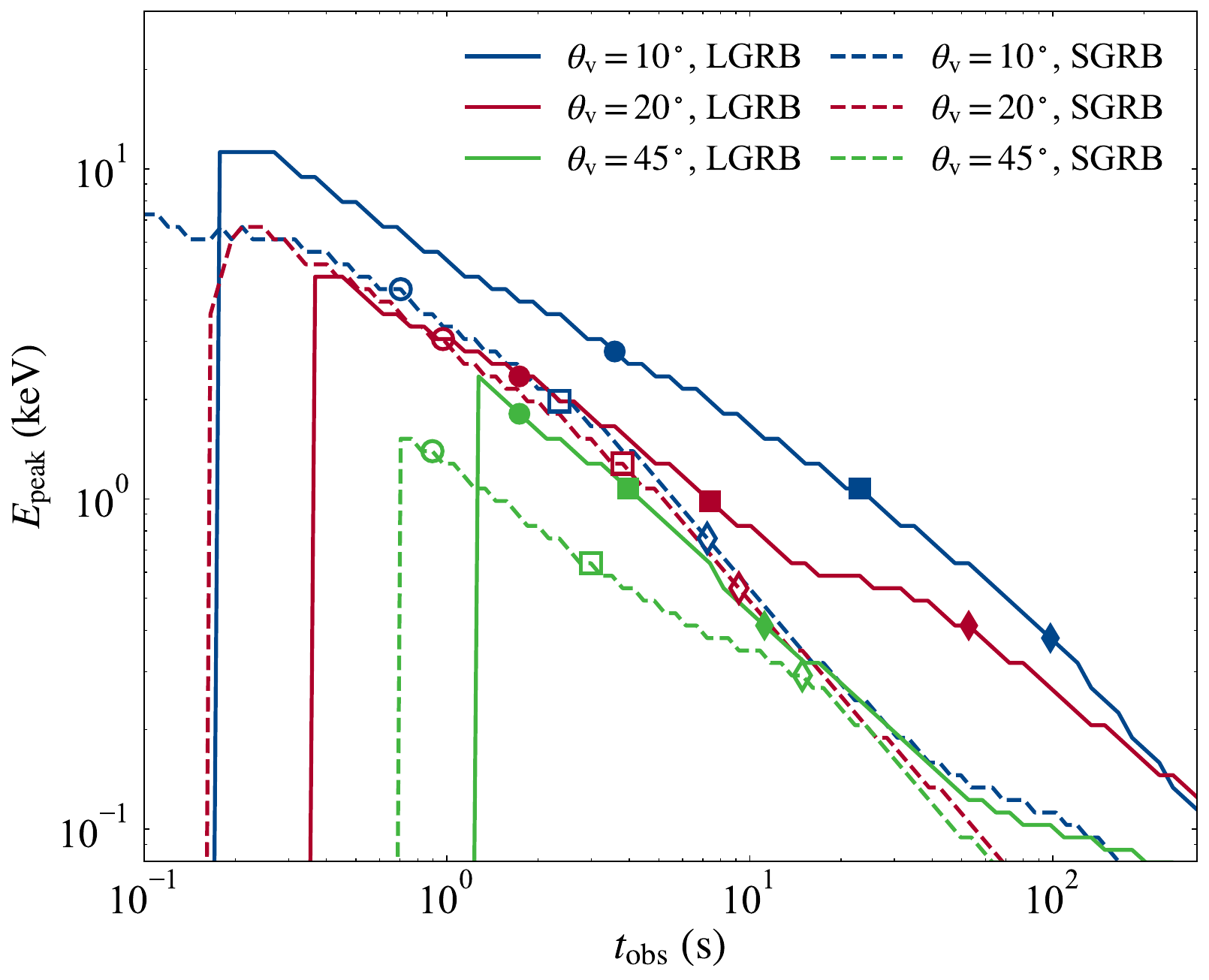}
    \caption{
    \textit{Left panel:} Time-resolved spectra of the canonical LGRB (solid lines) and SGRB (dashed lines) cocoon models at $\theta_{\rm v}=10^{\circ}$.
    The blue, yellow, and red curves correspond to $t_{\rm obs}=t_{5\%}$, $t_{50\%}$, and $t_{95\%}$, marking the beginning, midpoint, and end of the $T_{90}$ interval, respectively. All spectra are quasi-thermal. 
    \textit{Right panel:} Temporal evolutions of the peak energy for the canonical LGRB (solid lines) and SGRB (dashed lines) cocoon models.
    The blue, red, and green curves correspond to $\theta_{\rm v}=10^{\circ}$, $20^{\circ}$, and $45^{\circ}$, respectively. Circles, squares, and diamonds mark $t_{\rm obs}=t_{5\%}$, $t_{50\%}$, and $t_{95\%}$, respectively. Open and filled symbols denote the LGRB and SGRB cocoons.
    }
    \label{appfig:com}
\end{figure}

In this appendix, we discuss the main differences between LGRB and SGRB cocoons. 
As shown in Figure \ref{fig:Lc} $\&$ \ref{fig:avgmix}, the $T_{90}$-averaged spectra of SGRB cocoons are harder than those of LGRB cocoons in our previous results \citep{Zheng2026ApJ..1003L..19Z}, but only at small viewing angles. At larger viewing angles, the difference becomes smaller.

To understand this behavior, we compare the time-resolved spectra of the canonical model in two cocoons in the left panel of Figure \ref{appfig:com}. In both cases, the spectra are quasi-thermal and gradually soften with time. 
At the same fluence-normalized epochs within $T_{90}$, i.e., at $t_{5\%}$, $t_{50\%}$, and $t_{95\%}$, the SGRB cocoon has an $E_{\rm peak}$ larger by a factor of two. Here $t_{5\%}$, $t_{50\%}$, and $t_{95\%}$ denote the observer times when $5\%$, $50\%$, and $95\%$ of the total emitted energy in the EP-WXT band ($0.5$--$4\,{\rm keV}$) has been released, respectively. 
As a result, the SGRB cocoon at $\theta_{\rm v}=10^\circ$ appears harder in the EP WXT band than the LGRB cocoon.

However, this does not imply that the SGRB cocoon is intrinsically hotter at the same observer time. The right panel of Figure \ref{appfig:com} compares the evolution of $E_{\rm peak}$ in observer's time. At a fixed $t_{\rm obs}$, the SGRB cocoon has an $E_{\rm peak}$ comparable to, or even slightly lower than, that of the LGRB cocoon.
Because the SGRB cocoon emission has a shorter duration, the epochs  $t_{5\%}$, $t_{50\%}$, and $t_{95\%}$ correspond to earlier observer times, when the cocoon is hotter. In contrast, the LGRB cocoon with longer duration includes more late-time soft photons in its time-averaged spectrum, shifting the time-averaged spectral peak to lower energies.

This argument also explains why the SGRB cocoon is not always harder. At larger viewing angles, the duration of the LGRB cocoon is shorter and comparable to that of the SGRB cocoon. The time-averaging effect is then much weaker, so the SGRB spectrum is no longer harder. This viewing-angle dependence also affects the distribution in the Amati diagram (see Figure \ref{fig:Amati}). For LGRB cocoons, the averaged $E_{\rm peak}$ is nearly constant with viewing angle. For SGRB cocoons, however, $E_{\rm peak}$ shows a mild angular evolution, being higher at small viewing angles and lower at large viewing angles. This difference is partially caused by the anisotropic structure of the NSM ejecta, in contrast to the nearly spherical stellar envelope in the LGRB case. 
We verify this interpretation using an isotropic NSM ejecta model with the same ejecta mass $M_{\rm ej}$, in which the viewing-angle dependence of $E_{\rm peak}$ becomes weaker.


\begin{figure}
    \centering
    \includegraphics[width=0.5\linewidth]{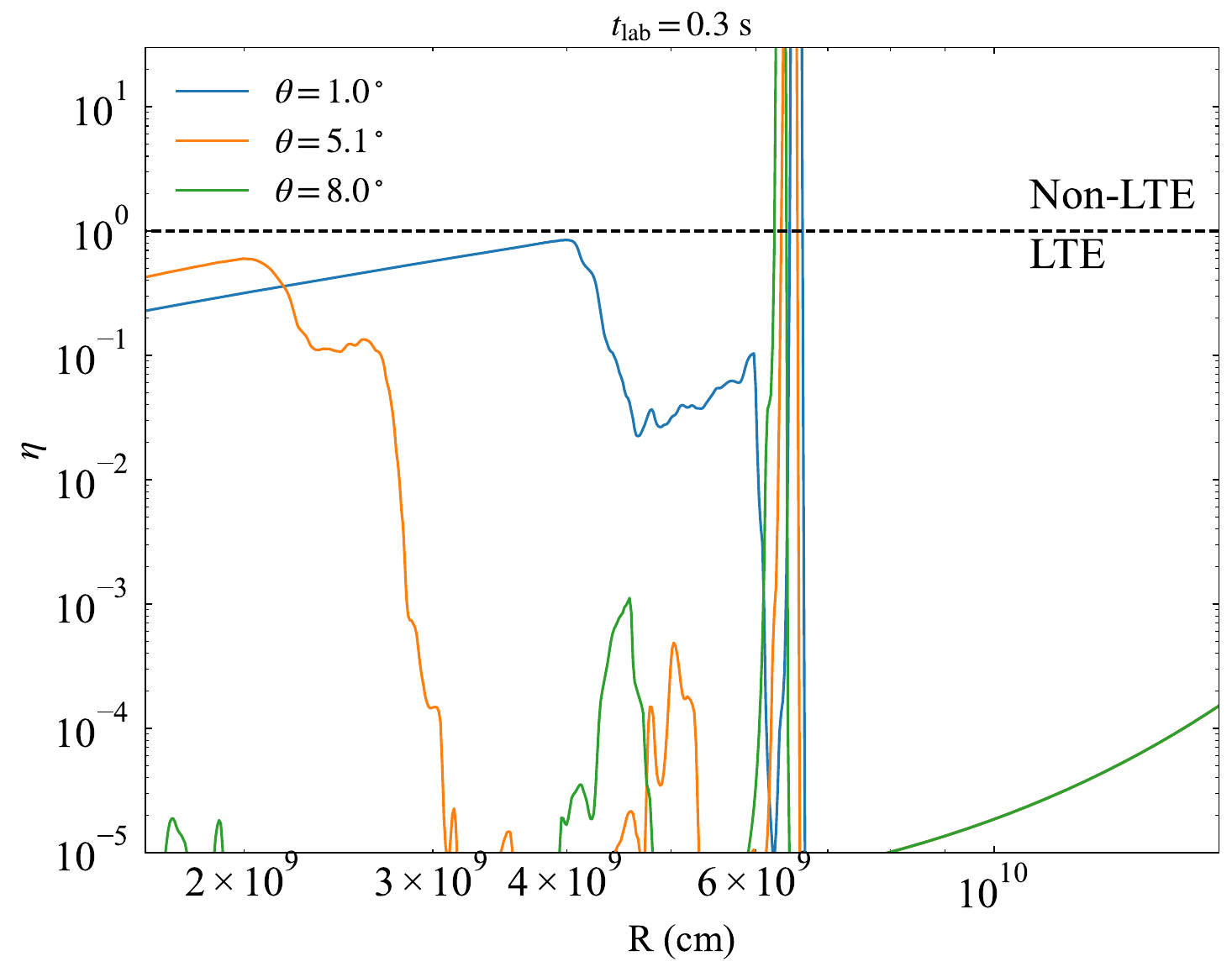}
    \caption{Thermalization parameter $\eta$ at the breakout time ($t_{\rm lab}=0.3$ s) for the canonical model. The blue, orange, and green curves correspond to $\theta=1.0^\circ$, $5.1^\circ$, and $8.0^\circ$, respectively. The dashed line marks $\eta=1$, separating the LTE and non-LTE regimes. The jet frontier is at $\sim6\times10^{9}$cm.
    }
    \label{fig:LTE}
\end{figure}

\section{Local thermodynamic equilibrium conditions}
\label{app:LTE}

In the main text, we assume that the radiation field in the jet–cocoon system is a blackbody in the comoving frame. This assumption may break down in relativistic outflows, where the low gas density can prevent sufficient photon production and thus inhibit thermalization within the dynamical expansion timescale.
We quantify the condition for local thermodynamic equilibrium (LTE) using
\begin{equation}
    \label{eq:LTE}
    \eta\equiv\frac{u'_{\rm int}}{\epsilon'_{\rm ff}t'_{\rm dyn}},
\end{equation}
where $u'_{\rm int}$ is the internal energy density and $t'_{\rm dyn}=r/(\Gamma\beta c)$ is the dynamical timescale. { The free--free emissivity for a fully ionized plasma is written as}
\begin{equation}
\epsilon'_{\rm ff}=
 1.4\times10^{-27} T'^{1/2}\frac{\langle Z^3\rangle}{\langle A^2\rangle} \left(\frac{\rho'}{m_p}\right)^2 {\rm erg\,cm^{-3}\,s^{-1}}
\end{equation}
{ The factor $\langle Z^3\rangle/\langle A^2\rangle$ characterizes the chemical composition of the ejecta. For a single ion species, the ion and electron number densities are $n'_i=\rho'/(A_i m_p)$ and  $n'_e=Z_in'_i$, giving}
\begin{equation}
\epsilon'_{\rm ff}\propto Z^2_in_in_e\propto\frac{Z_i^3}{A_i^2} \left(\frac{\rho'}{m_p}\right)^2.
\end{equation}
{ The $r$-process composition enhances free--free photon production because of the large ionic charge $Z_i$.  As a conservative estimate, even if the ejecta were dominated by first-peak $r$-process nuclei such as $^{88}{\rm Sr}$ ($Z=38$, $A=88$), the composition factor would be $Z_i^3/A_i^2\simeq7$.  Heavier $r$-process nuclei generally yield larger values. Therefore, we adopt $\langle Z^3\rangle/\langle A^2\rangle=7$ as a conservative value when evaluating the LTE condition.}

{
The parameter $\eta<1$ indicates that sufficient photons are produced to establish LTE during the shock-heating phase. Figure \ref{fig:LTE} shows $\eta$ for the canonical model ($Lc$) at the breakout time $t_{\rm lab}=0.3$ s. We evaluate the condition at breakout because most of the cocoon thermal energy is generated while the jet propagates through the ejecta. After breakout, the cocoon expands adiabatically without further heating. The cocoon material outside the jet core ($\theta\gtrsim5^\circ$) is safely in the LTE regime, while only the low-density shocked ISM ahead of the jet has $\eta>1$. }

This conclusion differs from that of \citet{Gottlieb2018CocoonSBO} because they considered a low-density, rapidly declining ``fast tail'' ($v_{\rm tail}\gtrsim0.3c$) outside the main NSM ejecta, with a density profile as steep as $\rho\propto r^{-10}$. 
They argued that the prompt gamma-ray emission of GRB 170817A was produced when the cocoon broke out of the fast tail.  Because the free–free production rate scales as $\epsilon'_{\rm ff}\propto n'^2_e$, the low-density breakout shell cannot produce enough photons to maintain LTE. The resulting photon-starved radiation has higher characteristic photon energies than a blackbody. 
This explains why shock-breakout emission can reach hard X-rays or gamma-rays. The cooling emission studied here is released from the denser bulk cocoon, where we verify that LTE is maintained. Its characteristic temperature is therefore much lower and naturally falls in the soft X-ray band.

\bibliography{reference}{}

\begin{thebibliography}{}
\expandafter\ifx\csname natexlab\endcsname\relax\def\natexlab#1{#1}\fi
\providecommand{\url}[1]{\href{#1}{#1}}
\providecommand{\dodoi}[1]{doi:~\href{http://doi.org/#1}{\nolinkurl{#1}}}
\providecommand{\doeprint}[1]{\href{http://ascl.net/#1}{\nolinkurl{http://ascl.net/#1}}}
\providecommand{\doarXiv}[1]{\href{https://arxiv.org/abs/#1}{\nolinkurl{https://arxiv.org/abs/#1}}}

\bibitem[{B.~P. {Abbott} {et~al.}(2017{\natexlab{a}}){Abbott}, {Abbott}, {Abbott}, {Acernese}, {Ackley}, {Adams}, {Adams}, {Addesso}, {Adhikari}, {Adya}, {Affeldt}, {Afrough}, {Agarwal}, {Agathos}, {Agatsuma}, {Aggarwal}, {Aguiar}, {Aiello}, {Ain}, {Ajith}, {Allen}, {Allen}, {Allocca}, {Altin}, {Amato}, {Ananyeva}, {Anderson}, {Anderson}, {Angelova}, {Antier}, {Appert}, {Arai}, {Araya}, {Areeda}, {Arnaud}, {Arun}, {Ascenzi}, {Ashton}, {Ast}, {Aston}, {Astone}, {Atallah}, {Aufmuth}, {Aulbert}, {AultONeal}, {Austin}, {Avila-Alvarez}, {Babak}, {Bacon}, {Bader}, {Bae}, {Bailes}, {Baker}, {Baldaccini}, {Ballardin}, {Ballmer}, {Banagiri}, {Barayoga}, {Barclay}, {Barish}, {Barker}, {Barkett}, {Barone}, {Barr}, {Barsotti}, {Barsuglia}, {Barta}, {Barthelmy}, {Bartlett}, {Bartos}, {Bassiri}, {Basti}, {Batch}, {Bawaj}, {Bayley}, {Bazzan}, {B{\'e}csy}, {Beer}, {Bejger}, {Belahcene}, {Bell}, {Berger}, {Bergmann}, {Bernuzzi}, {Bero}, {Berry}, {Bersanetti}, {Bertolini}, {Betzwieser}, {Bhagwat}, {Bhandare}, {Bilenko},
  {Billingsley}, {Billman}, {Birch}, {Birney}, {Birnholtz}, {Biscans}, {Biscoveanu}, {Bisht}, {Bitossi}, {Biwer}, {Bizouard}, {Blackburn}, {Blackman}, {Blair}, {Blair}, {Blair}, {Bloemen}, {Bock}, {Bode}, {Boer}, {Bogaert}, {Bohe}, {Bondu}, {Bonilla}, {Bonnand}, {Boom}, {Bork}, {Boschi}, {Bose}, {Bossie}, {Bouffanais}, {Bozzi}, {Bradaschia}, {Brady}, {Branchesi}, {Brau}, {Briant}, {Brillet}, {Brinkmann}, {Brisson}, {Brockill}, {Broida}, {Brooks}, {Brown}, {Brown}, {Brunett}, {Buchanan}, {Buikema}, {Bulik}, {Bulten}, {Buonanno}, {Buskulic}, {Buy}, {Byer}, {Cabero}, {Cadonati}, {Cagnoli}, {Cahillane}, {Calder{\'o}n Bustillo}, {Callister}, {Calloni}, {Camp}, {Canepa}, {Canizares}, {Cannon}, {Cao}, {Cao}, {Capano}, {Capocasa}, {Carbognani}, {Caride}, {Carney}, {Carullo}, {Casanueva Diaz}, {Casentini}, {Caudill}, {Cavagli{\`a}}, {Cavalier}, {Cavalieri}, {Cella}, {Cepeda}, {Cerd{\'a}-Dur{\'a}n}, {Cerretani}, {Cesarini}, {Chamberlin}, {Chan}, {Chao}, {Charlton}, {Chase}, {Chassande-Mottin}, {Chatterjee},
  {Chatziioannou}, {Cheeseboro}, {Chen}, {Chen}, {Chen}, {Cheng}, {Chia}, {Chincarini}, {Chiummo}, {Chmiel}, {Cho}, {Cho}, {Chow}, {Christensen}, {Chu}, {Chua}, \& {Chua}}]{GW170817PhRvL.119p1101A}
{Abbott}, B.~P., {Abbott}, R., {Abbott}, T.~D., {et~al.} 2017{\natexlab{a}}, \bibinfo{title}{{GW170817: Observation of Gravitational Waves from a Binary Neutron Star Inspiral},} \prl, 119, 161101, \dodoi{10.1103/PhysRevLett.119.161101}

\bibitem[{B.~P. {Abbott} {et~al.}(2017{\natexlab{b}}){Abbott}, {Abbott}, {Abbott}, {Acernese}, {Ackley}, {Adams}, {Adams}, {Addesso}, {Adhikari}, {Adya}, {Affeldt}, {Afrough}, {Agarwal}, {Agathos}, {Agatsuma}, {Aggarwal}, {Aguiar}, {Aiello}, {Ain}, {Ajith}, {Allen}, {Allen}, {Allocca}, {Aloy}, {Altin}, {Amato}, {Ananyeva}, {Anderson}, {Anderson}, {Angelova}, {Antier}, {Appert}, {Arai}, {Araya}, {Areeda}, {Arnaud}, {Arun}, {Ascenzi}, {Ashton}, {Ast}, {Aston}, {Astone}, {Atallah}, {Aufmuth}, {Aulbert}, {AultONeal}, {Austin}, {Avila-Alvarez}, {Babak}, {Bacon}, {Bader}, {Bae}, {Baker}, {Baldaccini}, {Ballardin}, {Ballmer}, {Banagiri}, {Barayoga}, {Barclay}, {Barish}, {Barker}, {Barkett}, {Barone}, {Barr}, {Barsotti}, {Barsuglia}, {Barta}, {Bartlett}, {Bartos}, {Bassiri}, {Basti}, {Batch}, {Bawaj}, {Bayley}, {Bazzan}, {B{\'e}csy}, {Beer}, {Bejger}, {Belahcene}, {Bell}, {Berger}, {Bergmann}, {Bero}, {Berry}, {Bersanetti}, {Bertolini}, {Betzwieser}, {Bhagwat}, {Bhandare}, {Bilenko}, {Billingsley}, {Billman},
  {Birch}, {Birney}, {Birnholtz}, {Biscans}, {Biscoveanu}, {Bisht}, {Bitossi}, {Biwer}, {Bizouard}, {Blackburn}, {Blackman}, {Blair}, {Blair}, {Blair}, {Bloemen}, {Bock}, {Bode}, {Boer}, {Bogaert}, {Bohe}, {Bondu}, {Bonilla}, {Bonnand}, {Boom}, {Bork}, {Boschi}, {Bose}, {Bossie}, {Bouffanais}, {Bozzi}, {Bradaschia}, {Brady}, {Branchesi}, {Brau}, {Briant}, {Brillet}, {Brinkmann}, {Brisson}, {Brockill}, {Broida}, {Brooks}, {Brown}, {Brown}, {Brunett}, {Buchanan}, {Buikema}, {Bulik}, {Bulten}, {Buonanno}, {Buskulic}, {Buy}, {Byer}, {Cabero}, {Cadonati}, {Cagnoli}, {Cahillane}, {Calder{\'o}n Bustillo}, {Callister}, {Calloni}, {Camp}, {Canepa}, {Canizares}, {Cannon}, {Cao}, {Cao}, {Capano}, {Capocasa}, {Carbognani}, {Caride}, {Carney}, {Casanueva Diaz}, {Casentini}, {Caudill}, {Cavagli{\`a}}, {Cavalier}, {Cavalieri}, {Cella}, {Cepeda}, {Cerd{\'a}-Dur{\'a}n}, {Cerretani}, {Cesarini}, {Chamberlin}, {Chan}, {Chao}, {Charlton}, {Chase}, {Chassande-Mottin}, {Chatterjee}, {Chatziioannou}, {Cheeseboro}, {Chen}, {Chen},
  {Chen}, {Cheng}, {Chia}, {Chincarini}, {Chiummo}, {Chmiel}, {Cho}, {Cho}, {Chow}, {Christensen}, {Chu}, {Chua}, {Chua}, {Chung}, {Chung}, \& {Ciani}}]{GW1708172017ApJ...848L..13A}
{Abbott}, B.~P., {Abbott}, R., {Abbott}, T.~D., {et~al.} 2017{\natexlab{b}}, \bibinfo{title}{{Gravitational Waves and Gamma-Rays from a Binary Neutron Star Merger: GW170817 and GRB 170817A},} \apjl, 848, L13, \dodoi{10.3847/2041-8213/aa920c}

\bibitem[{I. {Andreoni} {et~al.}(2017){Andreoni}, {Ackley}, {Cooke}, {Acharyya}, {Allison}, {Anderson}, {Ashley}, {Baade}, {Bailes}, {Bannister}, {Beardsley}, {Bessell}, {Bian}, {Bland}, {Boer}, {Booler}, {Brandeker}, {Brown}, {Buckley}, {Chang}, {Coward}, {Crawford}, {Crisp}, {Crosse}, {Cucchiara}, {Cup{\'a}k}, {de Gois}, {Deller}, {Devillepoix}, {Dobie}, {Elmer}, {Emrich}, {Farah}, {Farrell}, {Franzen}, {Gaensler}, {Galloway}, {Gendre}, {Giblin}, {Goobar}, {Green}, {Hancock}, {Hartig}, {Howell}, {Horsley}, {Hotan}, {Howie}, {Hu}, {Hu}, {James}, {Johnston}, {Johnston-Hollitt}, {Kaplan}, {Kasliwal}, {Keane}, {Kenney}, {Klotz}, {Lau}, {Laugier}, {Lenc}, {Li}, {Liang}, {Lidman}, {Luvaul}, {Lynch}, {Ma}, {Macpherson}, {Mao}, {McClelland}, {McCully}, {M{\"o}ller}, {Morales}, {Morris}, {Murphy}, {Noysena}, {Onken}, {Orange}, {Os{\l}owski}, {Pallot}, {Paxman}, {Potter}, {Pritchard}, {Raja}, {Ridden-Harper}, {Romero-Colmenero}, {Sadler}, {Sansom}, {Scalzo}, {Schmidt}, {Scott}, {Seghouani}, {Shang}, {Shannon},
  {Shao}, {Shara}, {Sharp}, {Sokolowski}, {Sollerman}, {Staff}, {Steele}, {Sun}, {Suntzeff}, {Tao}, {Tingay}, {Towner}, {Thierry}, {Trott}, {Tucker}, {V{\"a}is{\"a}nen}, {Krishnan}, {Walker}, {Wang}, {Wang}, {Wayth}, {Whiting}, {Williams}, {Williams}, {Wolf}, {Wu}, {Wu}, {Yang}, {Yuan}, {Zhang}, {Zhou}, \& {Zovaro}}]{Andreoni2017PASA...34...69A}
{Andreoni}, I., {Ackley}, K., {Cooke}, J., {et~al.} 2017, \bibinfo{title}{{Follow Up of GW170817 and Its Electromagnetic Counterpart by Australian-Led Observing Programmes},} \pasa, 34, e069, \dodoi{10.1017/pasa.2017.65}

\bibitem[{I. {Arcavi} {et~al.}(2017){Arcavi}, {Hosseinzadeh}, {Howell}, {McCully}, {Poznanski}, {Kasen}, {Barnes}, {Zaltzman}, {Vasylyev}, {Maoz}, \& {Valenti}}]{Arcavi2017Natur.551...64A}
{Arcavi}, I., {Hosseinzadeh}, G., {Howell}, D.~A., {et~al.} 2017, \bibinfo{title}{{Optical emission from a kilonova following a gravitational-wave-detected neutron-star merger},} \nat, 551, 64, \dodoi{10.1038/nature24291}

\bibitem[{S. {Banerjee} {et~al.}(2026){Banerjee}, {Hamidani}, {Kawaguchi}, \& {Tanaka}}]{Banerjee2026arXiv260120495B}
{Banerjee}, S., {Hamidani}, H., {Kawaguchi}, K., \& {Tanaka}, M. 2026, \bibinfo{title}{{Ultraviolet Signatures of Jet-Ejecta Interaction in Early Kilonovae: Prediction from Realistic Atomic Opacities},} arXiv e-prints, arXiv:2601.20495, \dodoi{10.48550/arXiv.2601.20495}

\bibitem[{E. {Berger}(2014){Berger}}]{Berger2014ARA&A..52...43B}
{Berger}, E. 2014, \bibinfo{title}{{Short-Duration Gamma-Ray Bursts},} \araa, 52, 43, \dodoi{10.1146/annurev-astro-081913-035926}

\bibitem[{A. {Chopra} {et~al.}(2026){Chopra}, {Ronchini}, {Banerjee}, {Branchesi}, {Ascenzi}, {Edvige Ravasio}, {Jonker}, \& {Levan}}]{Chopra2026arXiv260716420C}
{Chopra}, A., {Ronchini}, S., {Banerjee}, B., {et~al.} 2026, \bibinfo{title}{{Detectability of Gravitational-wave counterparts of EP-FXTs observed during the O4b LIGO-Virgo-KAGRA Observing Run},} arXiv e-prints, arXiv:2607.16420, \dodoi{10.48550/arXiv.2607.16420}

\bibitem[{D.~A. {Coulter} {et~al.}(2017){Coulter}, {Foley}, {Kilpatrick}, {Drout}, {Piro}, {Shappee}, {Siebert}, {Simon}, {Ulloa}, {Kasen}, {Madore}, {Murguia-Berthier}, {Pan}, {Prochaska}, {Ramirez-Ruiz}, {Rest}, \& {Rojas-Bravo}}]{Coulter2017Sci...358.1556C}
{Coulter}, D.~A., {Foley}, R.~J., {Kilpatrick}, C.~D., {et~al.} 2017, \bibinfo{title}{{Swope Supernova Survey 2017a (SSS17a), the optical counterpart to a gravitational wave source},} Science, 358, 1556, \dodoi{10.1126/science.aap9811}

\bibitem[{P.~S. {Cowperthwaite} {et~al.}(2017){Cowperthwaite}, {Berger}, {Villar}, {Metzger}, {Nicholl}, {Chornock}, {Blanchard}, {Fong}, {Margutti}, {Soares-Santos}, {Alexander}, {Allam}, {Annis}, {Brout}, {Brown}, {Butler}, {Chen}, {Diehl}, {Doctor}, {Drout}, {Eftekhari}, {Farr}, {Finley}, {Foley}, {Frieman}, {Fryer}, {Garc{\'\i}a-Bellido}, {Gill}, {Guillochon}, {Herner}, {Holz}, {Kasen}, {Kessler}, {Marriner}, {Matheson}, {Neilsen}, {Quataert}, {Palmese}, {Rest}, {Sako}, {Scolnic}, {Smith}, {Tucker}, {Williams}, {Balbinot}, {Carlin}, {Cook}, {Durret}, {Li}, {Lopes}, {Louren{\c{c}}o}, {Marshall}, {Medina}, {Muir}, {Mu{\~n}oz}, {Sauseda}, {Schlegel}, {Secco}, {Vivas}, {Wester}, {Zenteno}, {Zhang}, {Abbott}, {Banerji}, {Bechtol}, {Benoit-L{\'e}vy}, {Bertin}, {Buckley-Geer}, {Burke}, {Capozzi}, {Carnero Rosell}, {Carrasco Kind}, {Castander}, {Crocce}, {Cunha}, {D'Andrea}, {da Costa}, {Davis}, {DePoy}, {Desai}, {Dietrich}, {Drlica-Wagner}, {Eifler}, {Evrard}, {Fernandez}, {Flaugher}, {Fosalba}, {Gaztanaga},
  {Gerdes}, {Giannantonio}, {Goldstein}, {Gruen}, {Gruendl}, {Gutierrez}, {Honscheid}, {Jain}, {James}, {Jeltema}, {Johnson}, {Johnson}, {Kent}, {Krause}, {Kron}, {Kuehn}, {Nuropatkin}, {Lahav}, {Lima}, {Lin}, {Maia}, {March}, {Martini}, {McMahon}, {Menanteau}, {Miller}, {Miquel}, {Mohr}, {Neilsen}, {Nichol}, {Ogando}, {Plazas}, {Roe}, {Romer}, {Roodman}, {Rykoff}, {Sanchez}, {Scarpine}, {Schindler}, {Schubnell}, {Sevilla-Noarbe}, {Smith}, {Smith}, {Sobreira}, {Suchyta}, {Swanson}, {Tarle}, {Thomas}, {Thomas}, {Troxel}, {Vikram}, {Walker}, {Wechsler}, {Weller}, {Yanny}, \& {Zuntz}}]{Cowperthwaite2017ApJ...848L..17C}
{Cowperthwaite}, P.~S., {Berger}, E., {Villar}, V.~A., {et~al.} 2017, \bibinfo{title}{{The Electromagnetic Counterpart of the Binary Neutron Star Merger LIGO/Virgo GW170817. II. UV, Optical, and Near-infrared Light Curves and Comparison to Kilonova Models},} \apjl, 848, L17, \dodoi{10.3847/2041-8213/aa8fc7}

\bibitem[{M.~C. {D{\'\i}az} {et~al.}(2017){D{\'\i}az}, {Macri}, {Garcia Lambas}, {Mendes de Oliveira}, {Nilo Castell{\'o}n}, {Ribeiro}, {S{\'a}nchez}, {Schoenell}, {Abramo}, {Akras}, {Alcaniz}, {Artola}, {Beroiz}, {Bonoli}, {Cabral}, {Camuccio}, {Castillo}, {Chavushyan}, {Coelho}, {Colazo}, {Costa-Duarte}, {Cuevas Larenas}, {DePoy}, {Dom{\'\i}nguez Romero}, {Dultzin}, {Fern{\'a}ndez}, {Garc{\'\i}a}, {Girardini}, {Gon{\c{c}}alves}, {Gon{\c{c}}alves}, {Gurovich}, {Jim{\'e}nez-Teja}, {Kanaan}, {Lares}, {Lopes de Oliveira}, {L{\'o}pez-Cruz}, {Marshall}, {Melia}, {Molino}, {Padilla}, {Pe{\~n}uela}, {Placco}, {Qui{\~n}ones}, {Ram{\'\i}rez Rivera}, {Renzi}, {Riguccini}, {R{\'\i}os-L{\'o}pez}, {Rodriguez}, {Sampedro}, {Schneiter}, {Sodr{\'e}}, {Starck}, {Torres-Flores}, {Tornatore}, \& {Zadro{\.z}ny}}]{diaz2017ApJ...848L..29D}
{D{\'\i}az}, M.~C., {Macri}, L.~M., {Garcia Lambas}, D., {et~al.} 2017, \bibinfo{title}{{Observations of the First Electromagnetic Counterpart to a Gravitational-wave Source by the TOROS Collaboration},} \apjl, 848, L29, \dodoi{10.3847/2041-8213/aa9060}

\bibitem[{M.~R. {Drout} {et~al.}(2017){Drout}, {Piro}, {Shappee}, {Kilpatrick}, {Simon}, {Contreras}, {Coulter}, {Foley}, {Siebert}, {Morrell}, {Boutsia}, {Di Mille}, {Holoien}, {Kasen}, {Kollmeier}, {Madore}, {Monson}, {Murguia-Berthier}, {Pan}, {Prochaska}, {Ramirez-Ruiz}, {Rest}, {Adams}, {Alatalo}, {Ba{\~n}ados}, {Baughman}, {Beers}, {Bernstein}, {Bitsakis}, {Campillay}, {Hansen}, {Higgs}, {Ji}, {Maravelias}, {Marshall}, {Moni Bidin}, {Prieto}, {Rasmussen}, {Rojas-Bravo}, {Strom}, {Ulloa}, {Vargas-Gonz{\'a}lez}, {Wan}, \& {Whitten}}]{2017Sci...358.1570D}
{Drout}, M.~R., {Piro}, A.~L., {Shappee}, B.~J., {et~al.} 2017, \bibinfo{title}{{Light curves of the neutron star merger GW170817/SSS17a: Implications for r-process nucleosynthesis},} Science, 358, 1570, \dodoi{10.1126/science.aaq0049}

\bibitem[{P.~C. {Duffell} {et~al.}(2018){Duffell}, {Quataert}, {Kasen}, \& {Klion}}]{Duffell2018ApJ...866....3D}
{Duffell}, P.~C., {Quataert}, E., {Kasen}, D., \& {Klion}, H. 2018, \bibinfo{title}{{Jet Dynamics in Compact Object Mergers: GW170817 Likely Had a Successful Jet},} \apj, 866, 3, \dodoi{10.3847/1538-4357/aae084}

\bibitem[{D. {Eichler} {et~al.}(1989){Eichler}, {Livio}, {Piran}, \& {Schramm}}]{Eichler1989Natur.340..126E}
{Eichler}, D., {Livio}, M., {Piran}, T., \& {Schramm}, D.~N. 1989, \bibinfo{title}{{Nucleosynthesis, neutrino bursts and {\ensuremath{\gamma}}-rays from coalescing neutron stars},} \nat, 340, 126, \dodoi{10.1038/340126a0}

\bibitem[{S. {Fujibayashi} {et~al.}(2018){Fujibayashi}, {Kiuchi}, {Nishimura}, {Sekiguchi}, \& {Shibata}}]{Fujibayashi2018ApJ...860...64F}
{Fujibayashi}, S., {Kiuchi}, K., {Nishimura}, N., {Sekiguchi}, Y., \& {Shibata}, M. 2018, \bibinfo{title}{{Mass Ejection from the Remnant of a Binary Neutron Star Merger: Viscous-radiation Hydrodynamics Study},} \apj, 860, 64, \dodoi{10.3847/1538-4357/aabafd}

\bibitem[{A. {Goldstein} {et~al.}(2017){Goldstein}, {Veres}, {Burns}, {Briggs}, {Hamburg}, {Kocevski}, {Wilson-Hodge}, {Preece}, {Poolakkil}, {Roberts}, {Hui}, {Connaughton}, {Racusin}, {von Kienlin}, {Dal Canton}, {Christensen}, {Littenberg}, {Siellez}, {Blackburn}, {Broida}, {Bissaldi}, {Cleveland}, {Gibby}, {Giles}, {Kippen}, {McBreen}, {McEnery}, {Meegan}, {Paciesas}, \& {Stanbro}}]{GRB170817_2017ApJ...848L..14G}
{Goldstein}, A., {Veres}, P., {Burns}, E., {et~al.} 2017, \bibinfo{title}{{An Ordinary Short Gamma-Ray Burst with Extraordinary Implications: Fermi-GBM Detection of GRB 170817A},} \apjl, 848, L14, \dodoi{10.3847/2041-8213/aa8f41}

\bibitem[{O. {Gottlieb} {et~al.}(2018{\natexlab{a}}){Gottlieb}, {Nakar}, \& {Piran}}]{Gottlieb2018Cocoon}
{Gottlieb}, O., {Nakar}, E., \& {Piran}, T. 2018{\natexlab{a}}, \bibinfo{title}{{The cocoon emission - an electromagnetic counterpart to gravitational waves from neutron star mergers},} \mnras, 473, 576, \dodoi{10.1093/mnras/stx2357}

\bibitem[{O. {Gottlieb} {et~al.}(2018{\natexlab{b}}){Gottlieb}, {Nakar}, {Piran}, \& {Hotokezaka}}]{Gottlieb2018CocoonSBO}
{Gottlieb}, O., {Nakar}, E., {Piran}, T., \& {Hotokezaka}, K. 2018{\natexlab{b}}, \bibinfo{title}{{A cocoon shock breakout as the origin of the {\ensuremath{\gamma}}-ray emission in GW170817},} \mnras, 479, 588, \dodoi{10.1093/mnras/sty1462}

\bibitem[{E.~M. {Guti{\'e}rrez} {et~al.}(2025){Guti{\'e}rrez}, {Bhattacharya}, {Radice}, {Murase}, \& {Bernuzzi}}]{gutierrez2025PhRvD.111f3031G}
{Guti{\'e}rrez}, E.~M., {Bhattacharya}, M., {Radice}, D., {Murase}, K., \& {Bernuzzi}, S. 2025, \bibinfo{title}{{Cocoon shock breakout emission from binary neutron star mergers},} \prd, 111, 063031, \dodoi{10.1103/PhysRevD.111.063031}

\bibitem[{D. {Haggard} {et~al.}(2017){Haggard}, {Nynka}, {Ruan}, {Kalogera}, {Cenko}, {Evans}, \& {Kennea}}]{Haggard2017ApJ...848L..25H}
{Haggard}, D., {Nynka}, M., {Ruan}, J.~J., {et~al.} 2017, \bibinfo{title}{{A Deep Chandra X-Ray Study of Neutron Star Coalescence GW170817},} \apjl, 848, L25, \dodoi{10.3847/2041-8213/aa8ede}

\bibitem[{A. {Hajela} {et~al.}(2019){Hajela}, {Margutti}, {Alexander}, {Kathirgamaraju}, {Baldeschi}, {Guidorzi}, {Giannios}, {Fong}, {Wu}, {MacFadyen}, {Paggi}, {Berger}, {Blanchard}, {Chornock}, {Coppejans}, {Cowperthwaite}, {Eftekhari}, {Gomez}, {Hosseinzadeh}, {Laskar}, {Metzger}, {Nicholl}, {Paterson}, {Radice}, {Sironi}, {Terreran}, {Villar}, {Williams}, {Xie}, \& {Zrake}}]{Hajela2019ApJ...886L..17H}
{Hajela}, A., {Margutti}, R., {Alexander}, K.~D., {et~al.} 2019, \bibinfo{title}{{Two Years of Nonthermal Emission from the Binary Neutron Star Merger GW170817: Rapid Fading of the Jet Afterglow and First Constraints on the Kilonova Fastest Ejecta},} \apjl, 886, L17, \dodoi{10.3847/2041-8213/ab5226}

\bibitem[{H. {Hamidani} \& K. {Ioka}(2021){Hamidani} \& {Ioka}}]{Hamidani2021Propagation}
{Hamidani}, H., \& {Ioka}, K. 2021, \bibinfo{title}{{Jet propagation in expanding medium for gamma-ray bursts},} \mnras, 500, 627, \dodoi{10.1093/mnras/staa3276}

\bibitem[{H. {Hamidani} \& K. {Ioka}(2023){Hamidani} \& {Ioka}}]{Hamidani2023Cooling}
{Hamidani}, H., \& {Ioka}, K. 2023, \bibinfo{title}{{Cocoon cooling emission in neutron star mergers},} \mnras, 524, 4841, \dodoi{10.1093/mnras/stad1933}

\bibitem[{H. {Hamidani} {et~al.}(2025){Hamidani}, {Sato}, {Kashiyama}, {Tanaka}, {Ioka}, \& {Kimura}}]{Hamidani2025ApJ...986L...4H}
{Hamidani}, H., {Sato}, Y., {Kashiyama}, K., {et~al.} 2025, \bibinfo{title}{{EP240414a: A Gamma-Ray Burst Jet Weakened by an Extended Circumstellar Material},} \apjl, 986, L4, \dodoi{10.3847/2041-8213/add99d}

\bibitem[{K. {Hotokezaka} {et~al.}(2013){Hotokezaka}, {Kiuchi}, {Kyutoku}, {Okawa}, {Sekiguchi}, {Shibata}, \& {Taniguchi}}]{Hotokezaka2013PhRvD..87b4001H}
{Hotokezaka}, K., {Kiuchi}, K., {Kyutoku}, K., {et~al.} 2013, \bibinfo{title}{{Mass ejection from the merger of binary neutron stars},} \prd, 87, 024001, \dodoi{10.1103/PhysRevD.87.024001}

\bibitem[{K. {Hotokezaka} {et~al.}(2018){Hotokezaka}, {Kiuchi}, {Shibata}, {Nakar}, \& {Piran}}]{Hotokezaka2018ApJ...867...95H}
{Hotokezaka}, K., {Kiuchi}, K., {Shibata}, M., {Nakar}, E., \& {Piran}, T. 2018, \bibinfo{title}{{Synchrotron Radiation from the Fast Tail of Dynamical Ejecta of Neutron Star Mergers},} \apj, 867, 95, \dodoi{10.3847/1538-4357/aadf92}

\bibitem[{Z.-P. {Jin} {et~al.}(2016){Jin}, {Hotokezaka}, {Li}, {Tanaka}, {D'Avanzo}, {Fan}, {Covino}, {Wei}, \& {Piran}}]{Jin2016NatCo...712898J}
{Jin}, Z.-P., {Hotokezaka}, K., {Li}, X., {et~al.} 2016, \bibinfo{title}{{The Macronova in GRB 050709 and the GRB-macronova connection},} Nature Communications, 7, 12898, \dodoi{10.1038/ncomms12898}

\bibitem[{D. {Kasen} {et~al.}(2013){Kasen}, {Badnell}, \& {Barnes}}]{Kasen2013ApJ...774...25K}
{Kasen}, D., {Badnell}, N.~R., \& {Barnes}, J. 2013, \bibinfo{title}{{Opacities and Spectra of the r-process Ejecta from Neutron Star Mergers},} \apj, 774, 25, \dodoi{10.1088/0004-637X/774/1/25}

\bibitem[{D. {Kasen} {et~al.}(2017){Kasen}, {Metzger}, {Barnes}, {Quataert}, \& {Ramirez-Ruiz}}]{Kasen2017Natur.551...80K}
{Kasen}, D., {Metzger}, B., {Barnes}, J., {Quataert}, E., \& {Ramirez-Ruiz}, E. 2017, \bibinfo{title}{{Origin of the heavy elements in binary neutron-star mergers from a gravitational-wave event},} \nat, 551, 80, \dodoi{10.1038/nature24453}

\bibitem[{M.~M. {Kasliwal} {et~al.}(2017){Kasliwal}, {Nakar}, {Singer}, {Kaplan}, {Cook}, {Van Sistine}, {Lau}, {Fremling}, {Gottlieb}, {Jencson}, {Adams}, {Feindt}, {Hotokezaka}, {Ghosh}, {Perley}, {Yu}, {Piran}, {Allison}, {Anupama}, {Balasubramanian}, {Bannister}, {Bally}, {Barnes}, {Barway}, {Bellm}, {Bhalerao}, {Bhattacharya}, {Blagorodnova}, {Bloom}, {Brady}, {Cannella}, {Chatterjee}, {Cenko}, {Cobb}, {Copperwheat}, {Corsi}, {De}, {Dobie}, {Emery}, {Evans}, {Fox}, {Frail}, {Frohmaier}, {Goobar}, {Hallinan}, {Harrison}, {Helou}, {Hinderer}, {Ho}, {Horesh}, {Ip}, {Itoh}, {Kasen}, {Kim}, {Kuin}, {Kupfer}, {Lynch}, {Madsen}, {Mazzali}, {Miller}, {Mooley}, {Murphy}, {Ngeow}, {Nichols}, {Nissanke}, {Nugent}, {Ofek}, {Qi}, {Quimby}, {Rosswog}, {Rusu}, {Sadler}, {Schmidt}, {Sollerman}, {Steele}, {Williamson}, {Xu}, {Yan}, {Yatsu}, {Zhang}, \& {Zhao}}]{Kasliwal2017Sci...358.1559K}
{Kasliwal}, M.~M., {Nakar}, E., {Singer}, L.~P., {et~al.} 2017, \bibinfo{title}{{Illuminating gravitational waves: A concordant picture of photons from a neutron star merger},} Science, 358, 1559, \dodoi{10.1126/science.aap9455}

\bibitem[{G.~P. {Lamb} {et~al.}(2019){Lamb}, {Tanvir}, {Levan}, {de Ugarte Postigo}, {Kawaguchi}, {Corsi}, {Evans}, {Gompertz}, {Malesani}, {Page}, {Wiersema}, {Rosswog}, {Shibata}, {Tanaka}, {van der Horst}, {Cano}, {Fynbo}, {Fruchter}, {Greiner}, {Heintz}, {Higgins}, {Hjorth}, {Izzo}, {Jakobsson}, {Kann}, {O'Brien}, {Perley}, {Pian}, {Pugliese}, {Starling}, {Th{\"o}ne}, {Watson}, {Wijers}, \& {Xu}}]{Lamb2019ApJ...883...48L}
{Lamb}, G.~P., {Tanvir}, N.~R., {Levan}, A.~J., {et~al.} 2019, \bibinfo{title}{{Short GRB 160821B: A Reverse Shock, a Refreshed Shock, and a Well-sampled Kilonova},} \apj, 883, 48, \dodoi{10.3847/1538-4357/ab38bb}

\bibitem[{D. {Lazzati} {et~al.}(2017){Lazzati}, {L{\'o}pez-C{\'a}mara}, {Cantiello}, {Morsony}, {Perna}, \& {Workman}}]{Lazzati2017ApJ...848L...6L}
{Lazzati}, D., {L{\'o}pez-C{\'a}mara}, D., {Cantiello}, M., {et~al.} 2017, \bibinfo{title}{{Off-axis Prompt X-Ray Transients from the Cocoon of Short Gamma-Ray Bursts},} \apjl, 848, L6, \dodoi{10.3847/2041-8213/aa8f3d}

\bibitem[{A.~J. {Levan} {et~al.}(2024){Levan}, {Gompertz}, {Salafia}, {Bulla}, {Burns}, {Hotokezaka}, {Izzo}, {Lamb}, {Malesani}, {Oates}, {Ravasio}, {Rouco Escorial}, {Schneider}, {Sarin}, {Schulze}, {Tanvir}, {Ackley}, {Anderson}, {Brammer}, {Christensen}, {Dhillon}, {Evans}, {Fausnaugh}, {Fong}, {Fruchter}, {Fryer}, {Fynbo}, {Gaspari}, {Heintz}, {Hjorth}, {Kennea}, {Kennedy}, {Laskar}, {Leloudas}, {Mandel}, {Martin-Carrillo}, {Metzger}, {Nicholl}, {Nugent}, {Palmerio}, {Pugliese}, {Rastinejad}, {Rhodes}, {Rossi}, {Saccardi}, {Smartt}, {Stevance}, {Tohuvavohu}, {van der Horst}, {Vergani}, {Watson}, {Barclay}, {Bhirombhakdi}, {Breedt}, {Breeveld}, {Brown}, {Campana}, {Chrimes}, {D'Avanzo}, {D'Elia}, {De Pasquale}, {Dyer}, {Galloway}, {Garbutt}, {Green}, {Hartmann}, {Jakobsson}, {Kerry}, {Kouveliotou}, {Langeroodi}, {Le Floc'h}, {Leung}, {Littlefair}, {Munday}, {O'Brien}, {Parsons}, {Pelisoli}, {Sahman}, {Salvaterra}, {Sbarufatti}, {Steeghs}, {Tagliaferri}, {Th{\"o}ne}, {de Ugarte Postigo}, \&
  {Kann}}]{Levan2024Natur.626..737L}
{Levan}, A.~J., {Gompertz}, B.~P., {Salafia}, O.~S., {et~al.} 2024, \bibinfo{title}{{Heavy-element production in a compact object merger observed by JWST},} \nat, 626, 737, \dodoi{10.1038/s41586-023-06759-1}

\bibitem[{A.~J. {Levan} {et~al.}(2025){Levan}, {Jonker}, {Saccardi}, {Malesani}, {Tanvir}, {Izzo}, {Heintz}, {Mata S{\'a}nchez}, {Quirola-V{\'a}squez}, {Torres}, {Vergani}, {Schulze}, {Rossi}, {D'Avanzo}, {Gompertz}, {Martin-Carrillo}, {de Ugarte Postigo}, {Schneider}, {Yuan}, {Ling}, {Zhang}, {Mao}, {Liu}, {Sun}, {Xu}, {Zhu}, {Ag{\"u}{\'\i} Fern{\'a}ndez}, {Amati}, {Bauer}, {Campana}, {Carotenuto}, {Chrimes}, {van Dalen}, {D'Elia}, {Della Valle}, {De Pasquale}, {Dhillon}, {Galbany}, {Gaspari}, {Gianfagna}, {Gomboc}, {Habeeb}, {Hartmann}, {van Hoof}, {Hu}, {Jakobsson}, {Julakanti}, {Korth}, {Kouveliotou}, {Laskar}, {Littlefair}, {Maiorano}, {Mao}, {Melandri}, {Miller}, {Mukherjee}, {Oates}, {O'Brien}, {Palmerio}, {Parviainen}, {Pieterse}, {Piranomonte}, {Piro}, {Pugliese}, {Ravasio}, {Rayson}, {Salvaterra}, {S{\'a}nchez-Ram{\'\i}rez}, {Sarin}, {Shilling}, {Starling}, {Tagliaferri}, {Thakur}, {Th{\"o}ne}, {Wiersema}, {Worssam}, \& {Zafar}}]{Levan2025NatAs...9.1375L}
{Levan}, A.~J., {Jonker}, P.~G., {Saccardi}, A., {et~al.} 2025, \bibinfo{title}{{Fast X-ray transient EP240315A from a Lyman-continuum-leaking galaxy at z {\ensuremath{\approx}} 5},} Nature Astronomy, 9, 1375, \dodoi{10.1038/s41550-025-02612-9}

\bibitem[{A. {Li} {et~al.}(2026){Li}, {Wang}, {Passaleva}, {An}, {Zhang}, {Troja}, {Yin}, {Liu}, {Xiong}, {Xin}, {Shao}, {Yang}, {Sun}, {Xu}, {Yang}, {Ricci}, {Gao}, {Antier}, {Becerra}, {Cao}, {Castro-Tirado}, {Chen}, {Cheng}, {Chen}, {Cheng}, {D'Elia}, {De Pasquale}, {Dong}, {Elhosseiny}, {Eyles-Ferris}, {Gritsevich}, {Han}, {Hartmann}, {Hu}, {Hu}, {Jia}, {Kochiashvili}, {Lei}, {Levan}, {Li}, {Li}, {Li}, {Li}, {Ling}, {Liu}, {Lv}, {Malesani}, {O'Connor}, {Pan}, {Bhushan Pandey}, {Perez-Garcia}, {Pieterse}, {Pillas}, {Qiu}, {Saccardi}, {S{\'a}nchez-Ram{\'\i}rez}, {Tan}, {Tanasan}, {Tanvir}, {Vergani}, {Wang}, {Wang}, {Wu}, {Yi}, {Yusufjon}, {Zhang}, {Zhang}, {Zhang}, {Zhao}, {Zheng}, {Zheng}, {Zhou}, {Zhou}, {Cordier}, {Wei}, {Yuan}, {Zhang}, \& {Zhang}}]{Li2026_250704B}
{Li}, A., {Wang}, C.-W., {Passaleva}, N., {et~al.} 2026, \bibinfo{title}{{Minutes-long soft X-ray prompt emission from a compact object merger},} arXiv e-prints, arXiv:2601.14137, \dodoi{10.48550/arXiv.2601.14137}

\bibitem[{L.-X. {Li} \& B. {Paczy{\'n}ski}(1998){Li} \& {Paczy{\'n}ski}}]{Li1998ApJ...507L..59L}
{Li}, L.-X., \& {Paczy{\'n}ski}, B. 1998, \bibinfo{title}{{Transient Events from Neutron Star Mergers},} \apjl, 507, L59, \dodoi{10.1086/311680}

\bibitem[{W.~X. {Li} {et~al.}(2025){Li}, {Zhu}, {Zou}, {Geng}, {Liu}, {Wang}, {Li}, {Xu}, {Sun}, {Wang}, {Yu}, {Zhang}, {Wu}, {Yang}, {Filippenko}, {Liu}, {Yuan}, {Aguado}, {An}, {An}, {Buckley}, {Castro-Tirado}, {Fu}, {Fynbo}, {Howell}, {Hu}, {Jiang}, {Kumar}, {Mao}, {Maund}, {Liu}, {Mockler}, {Moskvitin}, {Andrews}, {Bom}, {Brink}, {Chatterjee}, {Chen}, {Cheng}, {Cooke}, {Dai}, {Du}, {Erasmus}, {Fang}, {Farah}, {Goranskij}, {Gritsevich}, {Gu}, {Guo}, {Hsiao}, {Hu}, {Hua}, {Jacobson-Gal{\'a}n}, {Jia}, {Jin}, {Kasliwal}, {Kilpatrick}, {Kumar}, {Lei}, {Li}, {Li}, {Li}, {Ling}, {Liu}, {Liu}, {Liu}, {L{\'o}pez-Oramas}, {Maslennikova}, {McCully}, {Monageng}, {Newsone}, {Padilla Gonzalez}, {Pan}, {Peng}, {Pignata}, {Poidevin}, {Potter}, {P{\'e}rez-Fournon}, {Santana-Silva}, {Santos}, {Song}, {Song}, {Spiridonova}, {Sun}, {Sun}, {Terreran}, {Wang}, {Wang}, {Wang}, {Wang}, {Wu}, {Xiang}, {Xiao}, {Xu}, {Xue}, {Yan}, {Yang}, {Yu}, {Zhang}, {Zhang}, {Zhang}, {Zhang}, {Zhang}, {Zheng}, \& {Zou}}]{Li2025arXiv250417034L}
{Li}, W.~X., {Zhu}, Z.~P., {Zou}, X.~Z., {et~al.} 2025, \bibinfo{title}{{An extremely soft and weak fast X-ray transient associated with a luminous supernova},} arXiv e-prints, arXiv:2504.17034, \dodoi{10.48550/arXiv.2504.17034}

\bibitem[{Y. {Liu} {et~al.}(2025){Liu}, {Sun}, {Xu}, {Svinkin}, {Delaunay}, {Tanvir}, {Gao}, {Zhang}, {Chen}, {Wu}, {Zhang}, {Yuan}, {An}, {Bruni}, {Frederiks}, {Ghirlanda}, {Hu}, {Li}, {Li}, {Li}, {Malesani}, {Piro}, {Raman}, {Ricci}, {Troja}, {Vergani}, {Wu}, {Yang}, {Zhang}, {Zhu}, {de Ugarte Postigo}, {Demin}, {Dobie}, {Fan}, {Fu}, {Fynbo}, {Geng}, {Gianfagna}, {Hu}, {Huang}, {Jiang}, {Jonker}, {Julakanti}, {Kennea}, {Kokomov}, {Kuulkers}, {Lei}, {Leung}, {Levan}, {Li}, {Li}, {Littlefair}, {Liu}, {Lysenko}, {Ma}, {Martin-Carrillo}, {O'Brien}, {Parsotan}, {Quirola-V{\'a}squez}, {Ridnaia}, {Ronchini}, {Rossi}, {Mata-S{\'a}nchez}, {Schneider}, {Shen}, {Thakur}, {Tohuvavohu}, {Torres}, {Tsvetkova}, {Ulanov}, {Wei}, {Xiao}, {Yin}, {Bai}, {Burwitz}, {Cai}, {Chen}, {Chen}, {Chen}, {Chen}, {Chen}, {Chen}, {Cheng}, {Cordier}, {Cui}, {Cui}, {Dai}, {Dai}, {Eder}, {Eyles-Ferris}, {Fan}, {Feldman}, {Feng}, {Feng}, {Friedrich}, {Gao}, {Gonzalez}, {Guan}, {Han}, {Han}, {Hou}, {Hu}, {Hu}, {Huang}, {Huo}, {Hutchinson},
  {Ji}, {Jia}, {Jia}, {Jiang}, {Jin}, {Jin}, {Jin}, {Keereman}, {Lerman}, {Li}, {Li}, {Li}, {Li}, {Li}, {Lian}, {Liang}, {Ling}, {Liu}, {Liu}, {Liu}, {Liu}, {Liu}, {Lu}, {L{\"u}}, {Luo}, {Ma}, {Ma}, {Mao}, {Mao}, {McHugh}, {Meidinger}, {Nandra}, {Osborne}, {Pan}, {Pan}, {Ravasio}, {Rau}, {Rea}, {Rehman}, {Sanders}, {Santovincenzo}, {Song}, {Su}, {Sun}, {Sun}, {Sun}, {Tan}, {Tang}, {Tao}, {Tong}, {Wang}, {Wang}, {Wang}, {Wang}, {Wang}, {Wang}, {Wang}, {Wang}, {Wang}, {Wei}, {Willingale}, {Xiong}, {Xu}, {Xu}, {Xu}, {Xu}, {Xu}, {Xue}, {Xue}, {Yan}, {Yang}, {Yang}, {Yang}, {Yang}, {Yu}, {Zhang}, {Zhang}, {Zhang}, {Zhang}, {Zhang}, {Zhang}, {Zhang}, {Zhang}, {Zhang}, {Zhao}, {Zhao}, {Zhao}, {Zhao}, {Zhou}, {Zhou}, {Zhu}, {Zhu}, \& {Zuo}}]{Liu2025NatAs...9..564L}
{Liu}, Y., {Sun}, H., {Xu}, D., {et~al.} 2025, \bibinfo{title}{{Soft X-ray prompt emission from the high-redshift gamma-ray burst EP240315a},} Nature Astronomy, 9, 564, \dodoi{10.1038/s41550-024-02449-8}

\bibitem[{S. {Makhathini} {et~al.}(2021){Makhathini}, {Mooley}, {Brightman}, {Hotokezaka}, {Nayana}, {Intema}, {Dobie}, {Lenc}, {Perley}, {Fremling}, {Mold{\`o}n}, {Lazzati}, {Kaplan}, {Balasubramanian}, {Brown}, {Carbone}, {Chandra}, {Corsi}, {Camilo}, {Deller}, {Frail}, {Murphy}, {Murphy}, {Nakar}, {Smirnov}, {Beswick}, {Fender}, {Hallinan}, {Heywood}, {Kasliwal}, {Lee}, {Lu}, {Rana}, {Perkins}, {White}, {J{\'o}zsa}, {Hugo}, \& {Kamphuis}}]{Makhathini2021ApJ...922..154M}
{Makhathini}, S., {Mooley}, K.~P., {Brightman}, M., {et~al.} 2021, \bibinfo{title}{{The Panchromatic Afterglow of GW170817: The Full Uniform Data Set, Modeling, Comparison with Previous Results, and Implications},} \apj, 922, 154, \dodoi{10.3847/1538-4357/ac1ffc}

\bibitem[{R. {Margutti} \& R. {Chornock}(2021){Margutti} \& {Chornock}}]{2021ARA&A..59..155M}
{Margutti}, R., \& {Chornock}, R. 2021, \bibinfo{title}{{First Multimessenger Observations of a Neutron Star Merger},} \araa, 59, 155, \dodoi{10.1146/annurev-astro-112420-030742}

\bibitem[{R. {Margutti} {et~al.}(2017){Margutti}, {Berger}, {Fong}, {Guidorzi}, {Alexander}, {Metzger}, {Blanchard}, {Cowperthwaite}, {Chornock}, {Eftekhari}, {Nicholl}, {Villar}, {Williams}, {Annis}, {Brown}, {Chen}, {Doctor}, {Frieman}, {Holz}, {Sako}, \& {Soares-Santos}}]{Margutti2017ApJ...848L..20M}
{Margutti}, R., {Berger}, E., {Fong}, W., {et~al.} 2017, \bibinfo{title}{{The Electromagnetic Counterpart of the Binary Neutron Star Merger LIGO/Virgo GW170817. V. Rising X-Ray Emission from an Off-axis Jet},} \apjl, 848, L20, \dodoi{10.3847/2041-8213/aa9057}

\bibitem[{R. {Margutti} {et~al.}(2018){Margutti}, {Alexander}, {Xie}, {Sironi}, {Metzger}, {Kathirgamaraju}, {Fong}, {Blanchard}, {Berger}, {MacFadyen}, {Giannios}, {Guidorzi}, {Hajela}, {Chornock}, {Cowperthwaite}, {Eftekhari}, {Nicholl}, {Villar}, {Williams}, \& {Zrake}}]{Margutti2018ApJ...856L..18M}
{Margutti}, R., {Alexander}, K.~D., {Xie}, X., {et~al.} 2018, \bibinfo{title}{{The Binary Neutron Star Event LIGO/Virgo GW170817 160 Days after Merger: Synchrotron Emission across the Electromagnetic Spectrum},} \apjl, 856, L18, \dodoi{10.3847/2041-8213/aab2ad}

\bibitem[{M. {Marinelli} \& J. {Green}(2025){Marinelli} \& {Green}}]{2025HST_WFC3IHB}
{Marinelli}, M., \& {Green}, J. 2025, {Wide Field Camera 3 Instrument Handbook, Version 18.0} (Baltimore, MD: Space Telescope Science Institute).
\newblock \url{https://hst-docs.stsci.edu/wfc3ihb}

\bibitem[{B.~D. {Metzger}(2020){Metzger}}]{Metzger2020LRR....23....1M}
{Metzger}, B.~D. 2020, \bibinfo{title}{{Kilonovae},} Living Reviews in Relativity, 23, 1, \dodoi{10.1007/s41114-019-0024-0}

\bibitem[{B.~D. {Metzger} \& E. {Berger}(2012){Metzger} \& {Berger}}]{Metzger2012ApJ...746...48M}
{Metzger}, B.~D., \& {Berger}, E. 2012, \bibinfo{title}{{What is the Most Promising Electromagnetic Counterpart of a Neutron Star Binary Merger?},} \apj, 746, 48, \dodoi{10.1088/0004-637X/746/1/48}

\bibitem[{B.~D. {Metzger} {et~al.}(2010){Metzger}, {Mart{\'\i}nez-Pinedo}, {Darbha}, {Quataert}, {Arcones}, {Kasen}, {Thomas}, {Nugent}, {Panov}, \& {Zinner}}]{Metzger2010MNRAS.406.2650M}
{Metzger}, B.~D., {Mart{\'\i}nez-Pinedo}, G., {Darbha}, S., {et~al.} 2010, \bibinfo{title}{{Electromagnetic counterparts of compact object mergers powered by the radioactive decay of r-process nuclei},} \mnras, 406, 2650, \dodoi{10.1111/j.1365-2966.2010.16864.x}

\bibitem[{A. {Mignone} {et~al.}(2007){Mignone}, {Bodo}, {Massaglia}, {Matsakos}, {Tesileanu}, {Zanni}, \& {Ferrari}}]{Mignone2007ApJS..170..228M}
{Mignone}, A., {Bodo}, G., {Massaglia}, S., {et~al.} 2007, \bibinfo{title}{{PLUTO: A Numerical Code for Computational Astrophysics},} \apjs, 170, 228, \dodoi{10.1086/513316}

\bibitem[{K.~P. {Mooley} {et~al.}(2022){Mooley}, {Anderson}, \& {Lu}}]{mooley2022_GW170817}
{Mooley}, K.~P., {Anderson}, J., \& {Lu}, W. 2022, \bibinfo{title}{{Optical superluminal motion measurement in the neutron-star merger GW170817},} \nat, 610, 273, \dodoi{10.1038/s41586-022-05145-7}

\bibitem[{K.~P. {Mooley} {et~al.}(2018){Mooley}, {Deller}, {Gottlieb}, {Nakar}, {Hallinan}, {Bourke}, {Frail}, {Horesh}, {Corsi}, \& {Hotokezaka}}]{Mooley2018Natur.561..355M}
{Mooley}, K.~P., {Deller}, A.~T., {Gottlieb}, O., {et~al.} 2018, \bibinfo{title}{{Superluminal motion of a relativistic jet in the neutron-star merger GW170817},} \nat, 561, 355, \dodoi{10.1038/s41586-018-0486-3}

\bibitem[{H. {Nagakura} {et~al.}(2014){Nagakura}, {Hotokezaka}, {Sekiguchi}, {Shibata}, \& {Ioka}}]{Nagakura2014ApJ...784L..28N}
{Nagakura}, H., {Hotokezaka}, K., {Sekiguchi}, Y., {Shibata}, M., \& {Ioka}, K. 2014, \bibinfo{title}{{Jet Collimation in the Ejecta of Double Neutron Star Mergers: A New Canonical Picture of Short Gamma-Ray Bursts},} \apjl, 784, L28, \dodoi{10.1088/2041-8205/784/2/L28}

\bibitem[{E. {Nakar}(2020){Nakar}}]{Nakar2020PhR...886....1N}
{Nakar}, E. 2020, \bibinfo{title}{{The electromagnetic counterparts of compact binary mergers},} \physrep, 886, 1, \dodoi{10.1016/j.physrep.2020.08.008}

\bibitem[{E. {Nakar} \& T. {Piran}(2017){Nakar} \& {Piran}}]{Nakar2017cocoon}
{Nakar}, E., \& {Piran}, T. 2017, \bibinfo{title}{{The Observable Signatures of GRB Cocoons},} \apj, 834, 28, \dodoi{10.3847/1538-4357/834/1/28}

\bibitem[{M. {Nicholl} {et~al.}(2017){Nicholl}, {Berger}, {Kasen}, {Metzger}, {Elias}, {Brice{\~n}o}, {Alexander}, {Blanchard}, {Chornock}, {Cowperthwaite}, {Eftekhari}, {Fong}, {Margutti}, {Villar}, {Williams}, {Brown}, {Annis}, {Bahramian}, {Brout}, {Brown}, {Chen}, {Clemens}, {Dennihy}, {Dunlap}, {Holz}, {Marchesini}, {Massaro}, {Moskowitz}, {Pelisoli}, {Rest}, {Ricci}, {Sako}, {Soares-Santos}, \& {Strader}}]{Nicholl2017ApJ...848L..18N}
{Nicholl}, M., {Berger}, E., {Kasen}, D., {et~al.} 2017, \bibinfo{title}{{The Electromagnetic Counterpart of the Binary Neutron Star Merger LIGO/Virgo GW170817. III. Optical and UV Spectra of a Blue Kilonova from Fast Polar Ejecta},} \apjl, 848, L18, \dodoi{10.3847/2041-8213/aa9029}

\bibitem[{M. {Nynka} {et~al.}(2018){Nynka}, {Ruan}, {Haggard}, \& {Evans}}]{Nynka2018ApJ...862L..19N}
{Nynka}, M., {Ruan}, J.~J., {Haggard}, D., \& {Evans}, P.~A. 2018, \bibinfo{title}{{Fading of the X-Ray Afterglow of Neutron Star Merger GW170817/GRB 170817A at 260 Days},} \apjl, 862, L19, \dodoi{10.3847/2041-8213/aad32d}

\bibitem[{B. {Paczynski}(1986){Paczynski}}]{Paczynski1986ApJ...308L..43P}
{Paczynski}, B. 1986, \bibinfo{title}{{Gamma-ray bursters at cosmological distances},} \apjl, 308, L43, \dodoi{10.1086/184740}

\bibitem[{M. {Pais} {et~al.}(2026){Pais}, {Ciolfi}, \& {Pavan}}]{Pais2026arXiv260629515P}
{Pais}, M., {Ciolfi}, R., \& {Pavan}, A. 2026, \bibinfo{title}{{Modelling the delayed shock-breakout emission following jet-launching binary neutron star mergers via relativistic magnetohydrodynamic simulations simulations},} arXiv e-prints, arXiv:2606.29515, \dodoi{10.48550/arXiv.2606.29515}

\bibitem[{A. {Perego} {et~al.}(2017){Perego}, {Radice}, \& {Bernuzzi}}]{PeregoAT2017gfo_2017ApJ...850L..37P}
{Perego}, A., {Radice}, D., \& {Bernuzzi}, S. 2017, \bibinfo{title}{{AT 2017gfo: An Anisotropic and Three-component Kilonova Counterpart of GW170817},} \apjl, 850, L37, \dodoi{10.3847/2041-8213/aa9ab9}

\bibitem[{E. {Pian} {et~al.}(2017){Pian}, {D'Avanzo}, {Benetti}, {Branchesi}, {Brocato}, {Campana}, {Cappellaro}, {Covino}, {D'Elia}, {Fynbo}, {Getman}, {Ghirlanda}, {Ghisellini}, {Grado}, {Greco}, {Hjorth}, {Kouveliotou}, {Levan}, {Limatola}, {Malesani}, {Mazzali}, {Melandri}, {M{\o}ller}, {Nicastro}, {Palazzi}, {Piranomonte}, {Rossi}, {Salafia}, {Selsing}, {Stratta}, {Tanaka}, {Tanvir}, {Tomasella}, {Watson}, {Yang}, {Amati}, {Antonelli}, {Ascenzi}, {Bernardini}, {Bo{\"e}r}, {Bufano}, {Bulgarelli}, {Capaccioli}, {Casella}, {Castro-Tirado}, {Chassande-Mottin}, {Ciolfi}, {Copperwheat}, {Dadina}, {De Cesare}, {di Paola}, {Fan}, {Gendre}, {Giuffrida}, {Giunta}, {Hunt}, {Israel}, {Jin}, {Kasliwal}, {Klose}, {Lisi}, {Longo}, {Maiorano}, {Mapelli}, {Masetti}, {Nava}, {Patricelli}, {Perley}, {Pescalli}, {Piran}, {Possenti}, {Pulone}, {Razzano}, {Salvaterra}, {Schipani}, {Spera}, {Stamerra}, {Stella}, {Tagliaferri}, {Testa}, {Troja}, {Turatto}, {Vergani}, \& {Vergani}}]{Pian2017Natur.551...67P}
{Pian}, E., {D'Avanzo}, P., {Benetti}, S., {et~al.} 2017, \bibinfo{title}{{Spectroscopic identification of r-process nucleosynthesis in a double neutron-star merger},} \nat, 551, 67, \dodoi{10.1038/nature24298}

\bibitem[{S. {Poolakkil} {et~al.}(2021){Poolakkil}, {Preece}, {Fletcher}, {Goldstein}, {Bhat}, {Bissaldi}, {Briggs}, {Burns}, {Cleveland}, {Giles}, {Hui}, {Kocevski}, {Lesage}, {Mailyan}, {Malacaria}, {Paciesas}, {Roberts}, {Veres}, {von Kienlin}, \& {Wilson-Hodge}}]{Poolakkil2021ApJ...913...60P}
{Poolakkil}, S., {Preece}, R., {Fletcher}, C., {et~al.} 2021, \bibinfo{title}{{The Fermi-GBM Gamma-Ray Burst Spectral Catalog: 10 yr of Data},} \apj, 913, 60, \dodoi{10.3847/1538-4357/abf24d}

\bibitem[{D. {Radice} {et~al.}(2018{\natexlab{a}}){Radice}, {Perego}, {Hotokezaka}, {Bernuzzi}, {Fromm}, \& {Roberts}}]{Radice2018Viscous}
{Radice}, D., {Perego}, A., {Hotokezaka}, K., {et~al.} 2018{\natexlab{a}}, \bibinfo{title}{{Viscous-dynamical Ejecta from Binary Neutron Star Mergers},} \apjl, 869, L35, \dodoi{10.3847/2041-8213/aaf053}

\bibitem[{D. {Radice} {et~al.}(2018{\natexlab{b}}){Radice}, {Perego}, {Hotokezaka}, {Fromm}, {Bernuzzi}, \& {Roberts}}]{Radice2018ApJ...869..130R}
{Radice}, D., {Perego}, A., {Hotokezaka}, K., {et~al.} 2018{\natexlab{b}}, \bibinfo{title}{{Binary Neutron Star Mergers: Mass Ejection, Electromagnetic Counterparts, and Nucleosynthesis},} \apj, 869, 130, \dodoi{10.3847/1538-4357/aaf054}

\bibitem[{J.~C. {Rastinejad} {et~al.}(2022){Rastinejad}, {Gompertz}, {Levan}, {Fong}, {Nicholl}, {Lamb}, {Malesani}, {Nugent}, {Oates}, {Tanvir}, {de Ugarte Postigo}, {Kilpatrick}, {Moore}, {Metzger}, {Ravasio}, {Rossi}, {Schroeder}, {Jencson}, {Sand}, {Smith}, {Ag{\"u}{\'\i} Fern{\'a}ndez}, {Berger}, {Blanchard}, {Chornock}, {Cobb}, {De Pasquale}, {Fynbo}, {Izzo}, {Kann}, {Laskar}, {Marini}, {Paterson}, {Escorial}, {Sears}, \& {Th{\"o}ne}}]{Rastinejad2022Natur.612..223R}
{Rastinejad}, J.~C., {Gompertz}, B.~P., {Levan}, A.~J., {et~al.} 2022, \bibinfo{title}{{A kilonova following a long-duration gamma-ray burst at 350 Mpc},} \nat, 612, 223, \dodoi{10.1038/s41586-022-05390-w}

\bibitem[{S. {Rosswog}(2007){Rosswog}}]{Rosswog2007MNRAS.376L..48R}
{Rosswog}, S. 2007, \bibinfo{title}{{Fallback accretion in the aftermath of a compact binary merger},} \mnras, 376, L48, \dodoi{10.1111/j.1745-3933.2007.00284.x}

\bibitem[{S. {Rosswog} {et~al.}(2025){Rosswog}, {Sarin}, {Nakar}, \& {Diener}}]{Rosswog2025MNRAS.538..907R}
{Rosswog}, S., {Sarin}, N., {Nakar}, E., \& {Diener}, P. 2025, \bibinfo{title}{{Fast dynamic ejecta in neutron star mergers},} \mnras, 538, 907, \dodoi{10.1093/mnras/staf324}

\bibitem[{G. {Ryan} {et~al.}(2020){Ryan}, {van Eerten}, {Piro}, \& {Troja}}]{Ryan2020ApJ...896..166R}
{Ryan}, G., {van Eerten}, H., {Piro}, L., \& {Troja}, E. 2020, \bibinfo{title}{{Gamma-Ray Burst Afterglows in the Multimessenger Era: Numerical Models and Closure Relations},} \apj, 896, 166, \dodoi{10.3847/1538-4357/ab93cf}

\bibitem[{B.~F. {Schutz}(2011){Schutz}}]{2011CQGra..28l5023S}
{Schutz}, B.~F. 2011, \bibinfo{title}{{Networks of gravitational wave detectors and three figures of merit},} Classical and Quantum Gravity, 28, 125023, \dodoi{10.1088/0264-9381/28/12/125023}

\bibitem[{B.~J. {Shappee} {et~al.}(2017){Shappee}, {Simon}, {Drout}, {Piro}, {Morrell}, {Prieto}, {Kasen}, {Holoien}, {Kollmeier}, {Kelson}, {Coulter}, {Foley}, {Kilpatrick}, {Siebert}, {Madore}, {Murguia-Berthier}, {Pan}, {Prochaska}, {Ramirez-Ruiz}, {Rest}, {Adams}, {Alatalo}, {Ba{\~n}ados}, {Baughman}, {Bernstein}, {Bitsakis}, {Boutsia}, {Bravo}, {Di Mille}, {Higgs}, {Ji}, {Maravelias}, {Marshall}, {Placco}, {Prieto}, \& {Wan}}]{Shappee2017Sci...358.1574S}
{Shappee}, B.~J., {Simon}, J.~D., {Drout}, M.~R., {et~al.} 2017, \bibinfo{title}{{Early spectra of the gravitational wave source GW170817: Evolution of a neutron star merger},} Science, 358, 1574, \dodoi{10.1126/science.aaq0186}

\bibitem[{M. {Shibata} \& K. {Hotokezaka}(2019){Shibata} \& {Hotokezaka}}]{Shibata2019Review}
{Shibata}, M., \& {Hotokezaka}, K. 2019, \bibinfo{title}{{Merger and Mass Ejection of Neutron Star Binaries},} Annual Review of Nuclear and Particle Science, 69, 41, \dodoi{10.1146/annurev-nucl-101918-023625}

\bibitem[{S.~J. {Smartt} {et~al.}(2017){Smartt}, {Chen}, {Jerkstrand}, {Coughlin}, {Kankare}, {Sim}, {Fraser}, {Inserra}, {Maguire}, {Chambers}, {Huber}, {Kr{\"u}hler}, {Leloudas}, {Magee}, {Shingles}, {Smith}, {Young}, {Tonry}, {Kotak}, {Gal-Yam}, {Lyman}, {Homan}, {Agliozzo}, {Anderson}, {Angus}, {Ashall}, {Barbarino}, {Bauer}, {Berton}, {Botticella}, {Bulla}, {Bulger}, {Cannizzaro}, {Cano}, {Cartier}, {Cikota}, {Clark}, {De Cia}, {Della Valle}, {Denneau}, {Dennefeld}, {Dessart}, {Dimitriadis}, {Elias-Rosa}, {Firth}, {Flewelling}, {Fl{\"o}rs}, {Franckowiak}, {Frohmaier}, {Galbany}, {Gonz{\'a}lez-Gait{\'a}n}, {Greiner}, {Gromadzki}, {Guelbenzu}, {Guti{\'e}rrez}, {Hamanowicz}, {Hanlon}, {Harmanen}, {Heintz}, {Heinze}, {Hernandez}, {Hodgkin}, {Hook}, {Izzo}, {James}, {Jonker}, {Kerzendorf}, {Klose}, {Kostrzewa-Rutkowska}, {Kowalski}, {Kromer}, {Kuncarayakti}, {Lawrence}, {Lowe}, {Magnier}, {Manulis}, {Martin-Carrillo}, {Mattila}, {McBrien}, {M{\"u}ller}, {Nordin}, {O'Neill}, {Onori}, {Palmerio}, {Pastorello},
  {Patat}, {Pignata}, {Podsiadlowski}, {Pumo}, {Prentice}, {Rau}, {Razza}, {Rest}, {Reynolds}, {Roy}, {Ruiter}, {Rybicki}, {Salmon}, {Schady}, {Schultz}, {Schweyer}, {Seitenzahl}, {Smith}, {Sollerman}, {Stalder}, {Stubbs}, {Sullivan}, {Szegedi}, {Taddia}, {Taubenberger}, {Terreran}, {van Soelen}, {Vos}, {Wainscoat}, {Walton}, {Waters}, {Weiland}, {Willman}, {Wiseman}, {Wright}, {Wyrzykowski}, \& {Yaron}}]{Smartt2017Natur.551...75S}
{Smartt}, S.~J., {Chen}, T.-W., {Jerkstrand}, A., {et~al.} 2017, \bibinfo{title}{{A kilonova as the electromagnetic counterpart to a gravitational-wave source},} \nat, 551, 75, \dodoi{10.1038/nature24303}

\bibitem[{H. {Sun} {et~al.}(2025){Sun}, {Li}, {Liu}, {Gao}, {Wang}, {Yuan}, {Zhang}, {Filippenko}, {Xu}, {An}, {Ai}, {Brink}, {Liu}, {Liu}, {Wang}, {Wu}, {Wu}, {Yang}, {Zhang}, {Zheng}, {Ahumada}, {Dai}, {Delaunay}, {Elias-Rosa}, {Benetti}, {Fu}, {Howell}, {Huang}, {Kasliwal}, {Karambelkar}, {Stein}, {Lei}, {Lian}, {Peng}, {Frederiks}, {Ridnaia}, {Svinkin}, {Wang}, {Wang}, {Wei}, {An}, {Andrews}, {Bai}, {Dai}, {Ehgamberdiev}, {Fan}, {Farah}, {Feng}, {Fynbo}, {Guo}, {Guo}, {Hu}, {Hu}, {Jiang}, {Jin}, {Li}, {Li}, {Li}, {Liang}, {Ling}, {Liu}, {Mao}, {McCully}, {Mirzaqulov}, {Newsome}, {Padilla Gonzalez}, {Pan}, {Terreran}, {Tinyanont}, {Wang}, {Wang}, {Wen}, {Xiang}, {Xue}, {Yang}, {Zhu}, {Cai}, {Castro-Tirado}, {Chen}, {Chen}, {Chen}, {Chen}, {Chen}, {Chen}, {Chen}, {Cheng}, {Cordier}, {Cui}, {Cui}, {Dai}, {Fan}, {Feng}, {Guan}, {Han}, {Hou}, {Hu}, {Huang}, {Huo}, {Jia}, {Jia}, {Jiang}, {Jin}, {Jin}, {Kuulkers}, {Li}, {Li}, {Li}, {Li}, {Li}, {Li}, {Li}, {Liu}, {Liu}, {Liu}, {Liu}, {Lu}, {Luo}, {Ma}, {Mao},
  {Nandra}, {O'Brien}, {Pan}, {Rau}, {Rea}, {Sanders}, {Song}, {Sun}, {Sun}, {Tan}, {Tang}, {Tao}, {Wang}, {Wang}, {Wang}, {Wang}, {Wang}, {Wang}, {Xiong}, {Xu}, {Xu}, {Xu}, {Xu}, {Xu}, {Xue}, {Xue}, {Yan}, {Yang}, {Yang}, {Yang}, {Zhang}, {Zhang}, {Zhang}, {Zhang}, {Zhang}, {Zhang}, {Zhang}, {Zhang}, {Zhang}, {Zhang}, {Zhao}, {Zhao}, {Zhao}, {Zhao}, {Zhou}, {Zhu}, {Zhu}, \& {Zou}}]{Sun2025NatAs...9.1073S}
{Sun}, H., {Li}, W.-X., {Liu}, L.-D., {et~al.} 2025, \bibinfo{title}{{A fast X-ray transient from a weak relativistic jet associated with a type Ic-BL supernova},} Nature Astronomy, 9, 1073, \dodoi{10.1038/s41550-025-02571-1}

\bibitem[{N.~R. {Tanvir} {et~al.}(2013){Tanvir}, {Levan}, {Fruchter}, {Hjorth}, {Hounsell}, {Wiersema}, \& {Tunnicliffe}}]{Tanvir2013Natur.500..547T}
{Tanvir}, N.~R., {Levan}, A.~J., {Fruchter}, A.~S., {et~al.} 2013, \bibinfo{title}{{A `kilonova' associated with the short-duration {\ensuremath{\gamma}}-ray burst GRB 130603B},} \nat, 500, 547, \dodoi{10.1038/nature12505}

\bibitem[{N.~R. {Tanvir} {et~al.}(2017){Tanvir}, {Levan}, {Gonz{\'a}lez-Fern{\'a}ndez}, {Korobkin}, {Mandel}, {Rosswog}, {Hjorth}, {D'Avanzo}, {Fruchter}, {Fryer}, {Kangas}, {Milvang-Jensen}, {Rosetti}, {Steeghs}, {Wollaeger}, {Cano}, {Copperwheat}, {Covino}, {D'Elia}, {de Ugarte Postigo}, {Evans}, {Even}, {Fairhurst}, {Figuera Jaimes}, {Fontes}, {Fujii}, {Fynbo}, {Gompertz}, {Greiner}, {Hodosan}, {Irwin}, {Jakobsson}, {J{\o}rgensen}, {Kann}, {Lyman}, {Malesani}, {McMahon}, {Melandri}, {O'Brien}, {Osborne}, {Palazzi}, {Perley}, {Pian}, {Piranomonte}, {Rabus}, {Rol}, {Rowlinson}, {Schulze}, {Sutton}, {Th{\"o}ne}, {Ulaczyk}, {Watson}, {Wiersema}, \& {Wijers}}]{Tanvir2017ApJ...848L..27T}
{Tanvir}, N.~R., {Levan}, A.~J., {Gonz{\'a}lez-Fern{\'a}ndez}, C., {et~al.} 2017, \bibinfo{title}{{The Emergence of a Lanthanide-rich Kilonova Following the Merger of Two Neutron Stars},} \apjl, 848, L27, \dodoi{10.3847/2041-8213/aa90b6}

\bibitem[{ {The LIGO Scientific Collaboration} {et~al.}(2026){The LIGO Scientific Collaboration}, {the Virgo Collaboration}, \& {the KAGRA Collaboration}}]{LIGO2026GWTC_5}
{The LIGO Scientific Collaboration}, {the Virgo Collaboration}, \& {the KAGRA Collaboration}. 2026, \bibinfo{title}{{GWTC-5.0: Population Properties of Merging Compact Binaries},} arXiv e-prints, arXiv:2605.27226, \dodoi{10.48550/arXiv.2605.27226}

\bibitem[{E. {Troja} {et~al.}(2017){Troja}, {Piro}, {van Eerten}, {Wollaeger}, {Im}, {Fox}, {Butler}, {Cenko}, {Sakamoto}, {Fryer}, {Ricci}, {Lien}, {Ryan}, {Korobkin}, {Lee}, {Burgess}, {Lee}, {Watson}, {Choi}, {Covino}, {D'Avanzo}, {Fontes}, {Gonz{\'a}lez}, {Khandrika}, {Kim}, {Kim}, {Lee}, {Lee}, {Kutyrev}, {Lim}, {S{\'a}nchez-Ram{\'\i}rez}, {Veilleux}, {Wieringa}, \& {Yoon}}]{Troja2017Natur.551...71T}
{Troja}, E., {Piro}, L., {van Eerten}, H., {et~al.} 2017, \bibinfo{title}{{The X-ray counterpart to the gravitational-wave event GW170817},} \nat, 551, 71, \dodoi{10.1038/nature24290}

\bibitem[{E. {Troja} {et~al.}(2019{\natexlab{a}}){Troja}, {Castro-Tirado}, {Becerra Gonz{\'a}lez}, {Hu}, {Ryan}, {Cenko}, {Ricci}, {Novara}, {S{\'a}nchez-R{\'a}mirez}, {Acosta-Pulido}, {Ackley}, {Caballero Garc{\'\i}a}, {Eikenberry}, {Guziy}, {Jeong}, {Lien}, {M{\'a}rquez}, {Pandey}, {Park}, {Sakamoto}, {Tello}, {Sokolov}, {Sokolov}, {Tiengo}, {Valeev}, {Zhang}, \& {Veilleux}}]{Troja2019MNRAS160821B}
{Troja}, E., {Castro-Tirado}, A.~J., {Becerra Gonz{\'a}lez}, J., {et~al.} 2019{\natexlab{a}}, \bibinfo{title}{{The afterglow and kilonova of the short GRB 160821B},} \mnras, 489, 2104, \dodoi{10.1093/mnras/stz2255}

\bibitem[{E. {Troja} {et~al.}(2019{\natexlab{b}}){Troja}, {van Eerten}, {Ryan}, {Ricci}, {Burgess}, {Wieringa}, {Piro}, {Cenko}, \& {Sakamoto}}]{Troja2019MNRAS170817}
{Troja}, E., {van Eerten}, H., {Ryan}, G., {et~al.} 2019{\natexlab{b}}, \bibinfo{title}{{A year in the life of GW 170817: the rise and fall of a structured jet from a binary neutron star merger},} \mnras, 489, 1919, \dodoi{10.1093/mnras/stz2248}

\bibitem[{V.~A. {Villar} {et~al.}(2017){Villar}, {Guillochon}, {Berger}, {Metzger}, {Cowperthwaite}, {Nicholl}, {Alexander}, {Blanchard}, {Chornock}, {Eftekhari}, {Fong}, {Margutti}, \& {Williams}}]{Villar2017ApJ...851L..21V}
{Villar}, V.~A., {Guillochon}, J., {Berger}, E., {et~al.} 2017, \bibinfo{title}{{The Combined Ultraviolet, Optical, and Near-infrared Light Curves of the Kilonova Associated with the Binary Neutron Star Merger GW170817: Unified Data Set, Analytic Models, and Physical Implications},} \apjl, 851, L21, \dodoi{10.3847/2041-8213/aa9c84}

\bibitem[{X.-Y. {Wang} \& Z.-Q. {Huang}(2018){Wang} \& {Huang}}]{Wang2018EarlyXray}
{Wang}, X.-Y., \& {Huang}, Z.-Q. 2018, \bibinfo{title}{{Early Soft X-Ray to UV Emission from Double Neutron Star Mergers: Implications from the Long-term Observations of GW170817},} \apjl, 853, L13, \dodoi{10.3847/2041-8213/aaa5fc}

\bibitem[{R. {Yamazaki} {et~al.}(2002){Yamazaki}, {Ioka}, \& {Nakamura}}]{Yamazaki2002ApJ...571L..31Y}
{Yamazaki}, R., {Ioka}, K., \& {Nakamura}, T. 2002, \bibinfo{title}{{X-Ray Flashes from Off-Axis Gamma-Ray Bursts},} \apjl, 571, L31, \dodoi{10.1086/341225}

\bibitem[{J. {Yang} {et~al.}(2022){Yang}, {Ai}, {Zhang}, {Zhang}, {Liu}, {Wang}, {Yang}, {Yin}, {Li}, \& {L{\"u}}}]{Yang2022Natur.612..232Y}
{Yang}, J., {Ai}, S., {Zhang}, B.-B., {et~al.} 2022, \bibinfo{title}{{A long-duration gamma-ray burst with a peculiar origin},} \nat, 612, 232, \dodoi{10.1038/s41586-022-05403-8}

\bibitem[{Y.-H. {Yang} {et~al.}(2024){Yang}, {Troja}, {O'Connor}, {Fryer}, {Im}, {Durbak}, {Paek}, {Ricci}, {Bom}, {Gillanders}, {Castro-Tirado}, {Peng}, {Dichiara}, {Ryan}, {van Eerten}, {Dai}, {Chang}, {Choi}, {De}, {Hu}, {Kilpatrick}, {Kutyrev}, {Jeong}, {Lee}, {Makler}, {Navarete}, \& {P{\'e}rez-Garc{\'\i}a}}]{Yang2024Natur.626..742Y}
{Yang}, Y.-H., {Troja}, E., {O'Connor}, B., {et~al.} 2024, \bibinfo{title}{{A lanthanide-rich kilonova in the aftermath of a long gamma-ray burst},} \nat, 626, 742, \dodoi{10.1038/s41586-023-06979-5}

\bibitem[{W. {Yuan} {et~al.}(2022){Yuan}, {Zhang}, {Chen}, \& {Ling}}]{Yuan2022hxga.book...86Y}
{Yuan}, W., {Zhang}, C., {Chen}, Y., \& {Ling}, Z. 2022, in Handbook of X-ray and Gamma-ray Astrophysics, ed. C.~{Bambi} \& A.~{Sangangelo} (Springer), 86, \dodoi{10.1007/978-981-16-4544-0_151-1}

\bibitem[{W. {Yuan} {et~al.}(2025){Yuan}, {Dai}, {Feng}, {Jin}, {Jonker}, {Kuulkers}, {Liu}, {Nandra}, {O'Brien}, {Piro}, {Rau}, {Rea}, {Sanders}, {Tao}, {Wang}, {Wu}, {Zhang}, {Zhang}, {Ai}, {Buchner}, {Bulbul}, {Chen}, {Chen}, {Chen}, {Chen}, {Coleiro}, {Coti Zelati}, {Dai}, {Fan}, {Fan}, {Friedrich}, {Gao}, {Ge}, {Ge}, {Geng}, {Ghirlanda}, {Gianfagna}, {Gou}, {Guillot}, {Hou}, {Hu}, {Huang}, {Ji}, {Jia}, {Komossa}, {Kong}, {Lan}, {Li}, {Li}, {Li}, {Li}, {Li}, {Li}, {Ling}, {Liu}, {Liu}, {Liu}, {Liu}, {Luo}, {Ma}, {Maggi}, {Maitra}, {Marino}, {Ng}, {Pan}, {Rukdee}, {Soria}, {Sun}, {Tam}, {Thakur}, {Tian}, {Troja}, {Wang}, {Wang}, {Wang}, {Wei}, {Wen}, {Wu}, {Wu}, {Xiao}, {Xu}, {Xu}, {Xu}, {Xu}, {Yang}, {You}, {Yu}, {Yu}, {Zhang}, {Zhang}, {Zhang}, {Zhang}, {Zhang}, {Zhang}, {Zhou}, \& {Zou}}]{Yuan2025SCPMA..6839501Y}
{Yuan}, W., {Dai}, L., {Feng}, H., {et~al.} 2025, \bibinfo{title}{{Science objectives of the Einstein Probe mission},} Science China Physics, Mechanics, and Astronomy, 68, 239501, \dodoi{10.1007/s11433-024-2600-3}

\bibitem[{B. {Zhang}(2013){Zhang}}]{Zhang2013ApJ...763L..22Z}
{Zhang}, B. 2013, \bibinfo{title}{{Early X-Ray and Optical Afterglow of Gravitational Wave Bursts from Mergers of Binary Neutron Stars},} \apjl, 763, L22, \dodoi{10.1088/2041-8205/763/1/L22}

\bibitem[{B.-B. {Zhang} {et~al.}(2018){Zhang}, {Zhang}, {Sun}, {Lei}, {Gao}, {Li}, {Shao}, {Zhao}, {Hu}, {L{\"u}}, {Wu}, {Fan}, {Wang}, {Castro-Tirado}, {Zhang}, {Yu}, {Cao}, \& {Liang}}]{Zhang2018GRB170817}
{Zhang}, B.-B., {Zhang}, B., {Sun}, H., {et~al.} 2018, \bibinfo{title}{{A peculiar low-luminosity short gamma-ray burst from a double neutron star merger progenitor},} Nature Communications, 9, 447, \dodoi{10.1038/s41467-018-02847-3}

\bibitem[{S.-N. {Zhang} {et~al.}(2025){Zhang}, {Santangelo}, {Xu}, {Feng}, {Lu}, {Chen}, {Ge}, {Nandra}, {Wu}, {Feroci}, {Hernanz}, {Liu}, {He}, {Wang}, {Jiang}, {Cui}, {Yang}, {Wang}, {Li}, {Li}, {Du}, {Liu}, {Meng}, {Wen}, {Zhang}, {Ma}, {Li}, {Li}, {Qi}, {Sun}, {Luo}, {Liu}, {Liu}, {Zhang}, {Luo}, {Zhu}, {Zhao}, {Sun}, {Yang}, {Wu}, {Jiang}, {Shi}, {Liu}, {Xu}, {Yang}, {Zhang}, {Han}, {Gao}, {Huo}, {Zhang}, {Wang}, {Zhao}, {Wang}, {Li}, {Bao}, {Liu}, {Wang}, {Wang}, {Wang}, {Wang}, {Wang}, {Ding}, {Sheng}, {Qiang}, {Yan}, {Liu}, {Wu}, {Liu}, {Chen}, {Zhang}, {Liu}, {Altmann}, {Bechteler}, {Burwitz}, {Fiorini}, {Friedrich}, {Meidinger}, {Strecker}, {Baldini}, {Bellazzini}, {Bonino}, {Frass{\`a}}, {Latronico}, {Maldera}, {Manfreda}, {Minuti}, {Pesce-Rollins}, {Sgr{\`o}}, {Tugliani}, {Pareschi}, {Basso}, {Sironi}, {Spiga}, {Tagliaferri}, {Tykhonov}, {Paltani}, {Bozzo}, {Tenzer}, {Bayer}, {Tuo}, {Liu}, {Zhang}, {Cai}, {Liu}, {Chen}, {Wang}, {He}, {Chen}, {Qiu}, {Zhang}, {Feng}, {Zhu}, {Zhou}, {Zheng}, {Song},
  {Wang}, {Jia}, {Jiang}, {Li}, {Zhao}, {Guan}, {Zhang}, {Li}, {Huang}, {Liao}, {You}, {Zhang}, {Wang}, {Wang}, {Ou}, {Hu}, {Shi}, {Cui}, {Jiang}, {Cheng}, {Li}, {Xu}, {Zane}, {Bambi}, {Bu}, {Dall'Osso}, {Rosa}, {Gou}, {Guillot}, {Ji}, {Li}, {Mao}, {Patruno}, {Stratta}, {Taverna}, {Tsygankov}, {Uttley}, {Watts}, {Wu}, {Xu}, {Yi}, {Zhang}, {Zhang}, {Zhao}, \& {Zhou}}]{Zhang2025eXTP..6819502Z}
{Zhang}, S.-N., {Santangelo}, A., {Xu}, Y., {et~al.} 2025, \bibinfo{title}{{The enhanced X-ray Timing and Polarimetry mission{\textemdash}eXTP for launch in 2030},} Science China Physics, Mechanics, and Astronomy, 68, 119502, \dodoi{10.1007/s11433-025-2786-6}

\bibitem[{J.-H. {Zheng} \& W. {Lu}(2026){Zheng} \& {Lu}}]{Zheng2026ApJ..1003L..19Z}
{Zheng}, J.-H., \& {Lu}, W. 2026, \bibinfo{title}{{Fast X-Ray Transients Produced by Off-axis Jet Cocoons from Long Gamma-Ray Bursts},} \apjl, 1003, L19, \dodoi{10.3847/2041-8213/ae67f2}

\bibitem[{J.-H. {Zheng} {et~al.}(2024){Zheng}, {Wang}, {Liu}, \& {Zhang}}]{Zheng2024ApJ...966..141Z}
{Zheng}, J.-H., {Wang}, X.-Y., {Liu}, R.-Y., \& {Zhang}, B. 2024, \bibinfo{title}{{A Narrow Uniform Core with a Wide Structured Wing: Modeling the TeV and Multiwavelength Afterglows of GRB 221009A},} \apj, 966, 141, \dodoi{10.3847/1538-4357/ad3949}

\bibitem[{J.-H. {Zheng} {et~al.}(2025){Zheng}, {Zhu}, {Lu}, \& {Zhang}}]{Zheng2025ApJ...985...21Z}
{Zheng}, J.-H., {Zhu}, J.-P., {Lu}, W., \& {Zhang}, B. 2025, \bibinfo{title}{{EP240414a: Off-axis View of a Jet-cocoon System from an Expanded Progenitor Star},} \apj, 985, 21, \dodoi{10.3847/1538-4357/adc993}

\end{thebibliography}
\bibliographystyle{aasjournal}

\end{document}